\PassOptionsToPackage{dvipsnames}{xcolor}
\documentclass[nofootinbib,longbibliography,preprintnumbers,amsmath,amssymb,12pt]{revtex4-2}
\usepackage{pgfpages}
\usepackage{footnote}
\usepackage{iftex}
\ifTUTeX

\else
\usepackage[T1]{fontenc}
\usepackage[utf8]{inputenc} 
\DeclareUnicodeCharacter{200B}{{\hskip 0pt}}
\fi
\usepackage[LGR,T1]{fontenc}
\usepackage{textgreek}
\usepackage{natbib}
\usepackage[english]{babel}
\usepackage[utf8]{inputenc}

\usepackage{enumitem}
\usepackage{braket}
\usepackage{footnote}
\usepackage{tablefootnote}
\usepackage[margin=1in]{geometry}
\usepackage{amsmath,amssymb,amsfonts,mathrsfs}
\usepackage{booktabs}
\usepackage{mathtools}
\usepackage{mathrsfs}

\usepackage[caption=false]{subfig}
\usepackage{bm}
\usepackage{dcolumn}
\usepackage{graphicx}   
\usepackage{latexsym}
\usepackage{cmbright}
\usepackage{pgfplots}
\usepackage[OT1]{fontenc}
\usepackage{kpfonts}
\usepackage[acronym]{glossaries}
\usepackage{physics}
\usepackage{siunitx}
\usepackage{comment}
\DeclareSIUnit{\persqrthz}{\ensuremath{\text{Hz}^{-1/2}}}
\usepackage[dvipsnames]{xcolor}
\usepackage[final]{hyperref}
\hypersetup{%
	colorlinks=true,
	citecolor=Blue,
}%

\newcommand{\myhyperref}[1]{\hyperref[#1]{\ref{#1}}}

\pgfplotsset{compat=newest}
\usepgfplotslibrary{external}
\tikzset{external/force remake}

\begin{document}
	
	\title{A class of charged and charged-Taub-NUT metrics in the presence of a massless scalar field and some of their astrophysical aspects}
	
	\author{Fatemeh Sadeghi}
	\email{fatemeh.sadeghi96@ph.iut.ac.ir}
	
	\author{Behrouz Mirza}
	\email{b.mirza@iut.ac.ir}
	
	\author{Marzieh Moradzadeh}
	\email{marziehmoradzadeh740@gmail.com}
	\affiliation{Department of Physics, Isfahan University of Technology, Isfahan 84156-83111, Iran}

	\begin{abstract}
		\vspace{5mm}
		\vspace{5mm}
		\begin{center}
			\textbf{Abstract}
		\end{center}
		We consider a class of exact solutions to the Einstein equations in the presence of a scalar field, recently introduced in \cite{Azizallahi2023, mirza2023class}, and derive their generalized form with dyonic charges using Harrison transformations. For specific parameter values, this class of metrics includes the charged Fisher-Janis-Newman-Winicour (FJNW) and  Zipoy-Voorhees (ZV) metrics. We then investigate the motion of neutral particles in the background of these metrics and derive the corresponding effective potential.
		
		Next, by applying Ehlers transformations, we introduce the NUT parameter into the Reissner-Nordstr\"{o}m (RN) metric in the presence of the scalar field. We also examine gravitational lensing, focusing on the effects of dyonic and NUT charges, as well as the scalar field, on the deflection angle of light. Finally, we explore the quasi-normal modes associated with this class of metrics.
	\end{abstract}
	
	
	\newacronym{e2e}{E2E}{End-To-End}
	\newacronym{inrep}{INREP}{Initial Noise REduction Pipeline}
	\newacronym{tdi}{TDI}{Time Delay Interferometry}
	\newacronym{ttl}{TTL}{Tilt-To-Length couplings}
	\newacronym{dfacs}{DFACS}{Drag-Free and Attitude Control System}
	\newacronym{ldc}{LDC}{LISA Data Challenge}
	\newacronym{lisa}{LISA}{the Laser Interferometer Space Antenna}
	\newacronym{emri}{EMRI}{Extreme Mass Ratio Inspiral}
	\newacronym{ifo}{IFO}{Interferometry System}
	\newacronym{grs}{GRS}{Gravitational Reference Sensor}
	\newacronym{tmdws}{TM-DWS}{Test-Mass Differential Wavefront Sensing}
	\newacronym{ldws}{LDWS}{Long-arm Differential Wavefront Sensing}
	\newacronym[	plural={MOSAs},
	first={Moving Optical Sub-Assembly},
	firstplural={Moving Optical Sub-Assemblies}
	]{mosa}{MOSA}{Moving Optical Sub-Assembly}
	\newacronym{siso}{SISO}{Single-Input Single-Output}
	\newacronym{mimo}{MIMO}{Multiple-Input Multiple-Output}
	\newacronym[plural=MBHB's, firstplural=Massive Black Holes Binaries (MBHB's)]{mbhb}{MBHB}{Massive Black Holes Binary}
	\newacronym{cmb}{CMB}{Cosmic Microwave Background}
	\newacronym{sgwb}{SGWB}{Stochastic Gravitational Waves Background}
	\newacronym{pta}{PTA}{Pulsar Timing Arrays}
	\newacronym{gw}{GW}{Gravitational Wave}
	\newacronym{snr}{SNR}{Signal-to-Noise Ratio}
	\newacronym{pbh}{PBH}{Primordial Black Holes}
	\newacronym{psd}{PSD}{Power Spectral Density}
	\newacronym{tcb}{TCB}{Barycentric Coordinate Time}
	\newacronym{bcrs}{BCRS}{Barycentric Celestial Reference System}
	\newacronym{lhs}{LHS}{Left-Hand Side}
	\newacronym{rhs}{RHS}{Right-Hand Side}
	\newacronym{mcmc}{MCMC}{Monte-Carlo Markov Chains}
	\newacronym{cs}{CS}{Cosmic Strings}
	\newacronym{ssb}{SSB}{Solar System Barycentric}
	\newacronym{oms}{OMS}{Optical Metrology System}
	\newacronym{dof}{DoF}{Degree of Freedom}
	\newacronym{eob}{EOB}{Effective One-Body}
	\newacronym{pn}{PN}{Post-Newtonian}
	\newacronym{cce}{CCE}{Cauchy-Characteristic Evolution}
	\newacronym{imr}{IMR}{Inspiral-Merger-Ringdown}
	\newacronym{scird}{SciRD}{LISA Science Requirement Document}
	
	
	%
	\maketitle
	
	\section{\label{Intro}Introduction}
	The study of the final state of an object undergoing gravitational collapse is one of the most important and intriguing topics in general relativity. Depending on the initial conditions and dynamics of the collapse, the final state of a gravitationally collapsing star can result in either the formation of a black hole \cite{hawking1972black, sachs1977general, adamek2016general} or a naked singularity \cite{penrose1973naked, christodoulou1994examples, harada2002physical, nakao2003does, szekeres2005naked, goswami2006quantum, joshi2009naked, ziaie2011naked, ortiz2015shadow, mohajan2017nature}. Unlike naked singularities, black holes are characterized by the presence of an event horizon. Notably, some singularities predicted by general relativity, such as the Big Bang singularity, are examples of naked singularities \cite{joshi2015story}.

A singularity becomes visible when the formation of an event horizon is delayed due to the internal dynamics of the collapse, allowing various phenomena to be observed by external observers. While classical general relativity permits the existence of such observable singularities, quantum gravity effects are expected to resolve or regularize these naked singularities. As such, naked singularities offer a unique laboratory for probing strong gravitational fields and exploring quantum gravity effects. In their vicinity, where gravitational forces are extreme and length and time scales approach the Planck scale, quantum effects are expected to play a significant role.

Combining quantum and gravitational effects is essential for a deeper understanding of spacetime under extreme conditions. During stellar collapse, it is plausible that classical singularities predicted by general relativity are resolved by quantum gravity effects. However, even if the final singularity is avoided, the collapse process still gives rise to regions of spacetime with extremely high mass density and curvature. Moreover, the collapse of a massive star may recreate physical conditions similar to those of the early universe. Unlike the singular event of the Big Bang, stellar collapse is a recurring phenomenon, occurring whenever a sufficiently massive star exhausts its nuclear fuel. If such extreme gravitational regions become observable, they would offer a rare opportunity to investigate the signatures of quantum gravity in the universe.

An important question is how a naked singularity can carry an electric charge. To address this, we first note that the density of matter and pressure within stars is typically very high, accompanied by an extremely strong gravitational field. Under such conditions, the mass distribution can be approximately proportional to the charge density, making it possible for a star to sustain both a significant electric charge and a strong internal electric field.

Due to the forces acting on a charged particle within a star, we expect that individual charged particles will tend to be pushed outward by the Coulomb repulsion. The repeated occurrence of this process leads to an imbalance of forces within the star. Eventually, the gravitational attraction overcomes both the Coulomb force and the internal fluid pressure. Just before all the charged particles escape, the system undergoes gravitational collapse, resulting in the formation of a charged singularity. For a more detailed discussion on the influence of electric fields and charge in gravitational systems, see \cite{hawking1971gravitationally, ray2003electrically}.

In addition to electric and magnetic charges, spacetime itself can possess a gravomagnetic monopole. A metric incorporating such a feature was first introduced by Taub as a homogeneous vacuum cosmological model \cite{taub1951empty}, and later generalized by Newman, Unti, and Tamburino (NUT) as an extension of the Schwarzschild solution \cite{newman1963empty, kramer1980exact}. The resulting Taub-NUT metrics exhibit a range of intriguing and unconventional properties, leading to various physical interpretations. For early discussions, see \cite{misner1963flatter, misner1967contribution, bonnor1969new, manko2005physical}. For a more comprehensive treatment of the NUT and Taub-NUT spacetimes, including their thermodynamics, gravitational lensing, and quasinormal modes, refer to \cite{gonzalez2018new, hennigar2019thermodynamics, bordo2019first, bordo2019thermodynamics, bordo2019misner, zhang2021nut, arratia2021hairy, barrientos2022gravitational, durka2022first, cano2022quasinormal, derekeh2024class, kachi2025class, barrientos2025rotating, jafarzade2025modelling}.

The nature of a singularity is determined by the properties of the collapsing matter and the initial conditions of the system. For example, the gravitational collapse of a perfectly spherical, charged body leads to the formation of a Reissner-Nordstr\"{o}m (RN) black hole—the simplest known example of a charged black hole. The influence of electric charge also extends to other spacetime geometries. One notable case is the Zipoy-Voorhees (ZV) metric \cite{darmois1927memorial, erez1959gravitational, zipoy1966topology, voorhees1970static}, which describes a static, axially symmetric spacetime and has been extensively studied, including in its higher-dimensional generalizations \cite{hajibarat2022gamma}. Furthermore, a charged version of the ZV metric ($\gamma$-metric) has been explored in \cite{gurtug2022charged}.

Recently, the ZV metric has been proposed as a viable model for describing the compact object at the center of our galaxy, as it successfully reproduces many of the observed features of this region \cite{destounis2023geodesics, lora2023textit}. Another significant spacetime is the Fisher-Janis- Newman-Winicour (FJNW) metric \cite{fisher1999scalar, janis1968reality, wyman1981static}, which represents a static, axisymmetric solution to the Einstein equations in the presence of a massless scalar field. The charged extension of the FJNW metric was first derived in \cite{penney1969generalization}. Also, Bogush and Gal'tsov \cite{bogush2020generation} demonstrated that previously proposed rotating FJNW solutions were incorrect, as they did not satisfy the Einstein field equations. They also proposed a rotating FJNW solution for specific values of the parameters and with a phantom scalar field. Later, in \cite{mirza2023class}, for the first time a general physical rotating solution for FJNW has been found that is valid for all range of parameters with a real scalar field. Many studies have examined the FJNW spacetime and the Einstein field equations minimally coupled to a scalar field \cite{penney1968axially, janis1969comments, xanthopoulos1992kantowski, bogush2020generation, Azizallahi2023, barrientos2024revisiting, mirza2023class, mirza2024exact, derekeh2024class, kachi2025class, jafarzade2025modelling, barrientos2025rotating}. 

In \cite{ernst1968new, ernst1968new2}, Ernst reformulated the Einstein-Maxwell theory by introducing complex potential functions, leading to a more compact and symmetric representation of the field equations. These equations exhibit a rich symmetry structure and remain invariant under certain solution-generating techniques, notably the Harrison \cite{harrison1968new} and Ehlers \cite{ehlers1958konstruktionen} transformations. The Harrison transformation introduces a dyonic charge into the spacetime, while the Ehlers transformation incorporates the NUT parameter. These transformations can also be extended to the Einstein theory coupled to a scalar field, as explored in \cite{erics1977stationary, astorino2013embedding, astorino2015stationary}. Recently, by using the Ernst scheme a new rotating solution for the vacuum Einstein's equations was derived in \cite{barrientos2025new}.

In this paper, we investigate a class of exact solutions to the Einstein equations recently proposed in \cite{Azizallahi2023, mirza2023class, mirza2024exact}. By applying the Harrison and Ehlers transformations, we construct a new family of static, charged, Taub-NUT solutions that are asymptotically flat and include a massless scalar field. We then analyze the astrophysical properties of these newly derived charged Taub-NUT spacetimes.

Geodesic motion plays a fundamental role in the study of compact astrophysical objects, offering crucial insights into the underlying geometry of spacetime. Among the various types of geodesics, circular orbits are of particular interest due to their relevance in observational astrophysics and their connection to stability analyses \cite{chandrasekhar1983mathematical, cardoso2009geodesic}. In Sec. \ref{4a}, we analyze the geodesic structure of the newly proposed class of metrics and identify the allowable range for one of the free parameters based on the conditions for physically viable circular motion.

One of the most intriguing topics in astrophysics is the study of the motion of neutral or charged, massless or massive particles in the vicinity of compact objects. The trajectories of such particles are shaped by the characteristics of the gravitational source, offering valuable information about the nature of the underlying spacetime \cite{sharp1979geodesics, frolov2012black}. By examining these trajectories, one can gain insights into both the physical and geometric properties of the compact object. In particular, the motion of neutral and charged particles serves as a powerful tool for probing the structure of various spacetimes \cite{pugliese2013equatorial, kovavr2014electrically, zakharov2014constraints, stuchlik2015circular, toshmatov2015particle, kovavr2016charged, tursunov2016circular, stuchlik2016acceleration, beheshti2016marginally}.

Gravitational lensing, which examines the deflection of light by massive objects, is one of the most powerful tools for exploring both known and unknown aspects of the universe. Various methods have been developed to study gravitational lensing, providing deep insights into the geometry and structure of spacetime. For a comprehensive overview of these techniques, see, for example, \cite{schneider1992gravitational, wambsganss1998gravitational, narayan1999gravitational, schneider2006gravitational, straumann2012general}.

In astronomy, weak gravitational lensing is the most commonly observed phenomenon. However, the situation is quite different for light bending near compact objects, where the effects can be much more pronounced. For studies specifically addressing gravitational lensing around moving and rotating astrophysical objects, see \cite{ibanez1983gravitational, birkinshaw1983test, sereno2003gravitational, heyrovsky2005velocity, schaefer2006weak, gibbons2008applications}.

In this paper, we calculate the deflection angle of light for the charged spacetimes in the presence of a scalar field.

Another approach to understanding the properties of compact astrophysical objects is by studying their interactions with the surrounding environment. The perturbations of such objects give rise to the emission of gravitational waves, which can be analyzed through their corresponding quasi-normal modes (QNMs). QNMs are characterized by damped oscillations that evolve over time, providing valuable information about the object's internal structure and the nature of its spacetime. For more details on QNMs and their calculation methods, refer to \cite{nollert1996significance, nollert1999quasinormal, kokkotas1999quasi, dreyer2003quasinormal, berti2009quasinormal, denef2010black, konoplya2011quasinormal, flachi2013quasinormal, matyjasek2017quasinormal}. In Sec. \ref{6a}, we employ the light ring method \cite{ferrari1984new, ferrari1984oscillations, mashhoon1985stability} to derive the QNMs.

The paper is organized as follows. In Sec. \ref{2}, we derive the charged versions of the metrics introduced in \cite{Azizallahi2023} using Harrison transformations. For specific choices of parameters, these metrics reduce to the charged FJNW and charged ZV metrics. We then analyze geodesic motion to constrain the permissible range of the metric parameters. Additionally, we investigate the dynamics of neutral particles in the background of these spacetimes.

In Sec. \ref{CTNS}, we apply both the Ehlers and Harrison transformations to construct a class of charged Taub-NUT metrics in the presence of a scalar field. Section \ref{5} is devoted to the study of gravitational lensing, where we derive the deflection angle of light in scenarios with and without charge, the NUT parameter, and the scalar field. In Sec. \ref{6a}, we compute the quasi-normal modes (QNMs) for specific parameter limits. Finally, Sec. \ref{6} summarizes our findings and presents the conclusions.

In this article, we adopt natural units by setting $ 8\,\pi\,G=c=1 $.
\section{A class of static, charged metrics in the presence of a massless scalar field}\label{2}
In this section, we derive a class of static, axially symmetric solutions to the Einstein-Maxwell equations in the presence of a massless scalar field. To do so, we will employ Harrison transformations. The Einstein-Maxwell action, in the presence of a massless scalar field $ \varphi(r) $, can be written as follows
\begin{equation}\label{100_label}
	S = \int d^4 x \, \sqrt{-g} \, [R - \partial_{\mu} \varphi \, \partial^{\mu} \varphi -  F_{\mu \nu} \, F^{\mu \nu}].
\end{equation}
The field equations corresponding to Eq. \eqref{100_label} are:
\begin{equation}\label{101_label}
	R_{\mu \nu} = \partial_{\mu} \varphi(r) \, \partial_{\nu} \varphi(r) + T_{\mu \nu}^{EM},
\end{equation}
\begin{equation}\label{101b_label}
	\Box \varphi(r) = 0,
\end{equation}
where $ T_{\mu \nu}^{EM} $ is given by
\begin{equation}\label{715a_label}
	T_{\mu \nu}^{EM} = 2 \, F_\mu^\alpha \, F_{\nu \alpha} - \frac{1}{2} \, g_{\mu \nu} \, F_{\gamma \sigma} \, F^{\gamma \sigma},
\end{equation}
where, $  F^{\mu \nu} = \partial^\mu \, A^\nu -  \partial^\nu \, A^\mu $ is the electromagnetic field tensor. In the absence of the electromagnetic energy-momentum tensor ($T_{\mu\nu}^{EM} =0$), a class of static solutions to the equations of motion \eqref{101_label} and \eqref{101b_label} is given by the following metrics
\cite{Azizallahi2023}
\begin{equation}\label{last1_label}
	d s^2 = - f^\gamma \, dt^2 + f^\mu \, k^\nu \, \big( \frac{d r^2}{f} + r^2 d \theta^2 \big) + r^2 f^{1 - \gamma} \sin^2 \theta \, d \phi^2,
\end{equation}
where,
\begin{equation}\label{last2_label}
	f(r) = 1 - \frac{2\,m}{r}, \qquad k(r, \theta) = 1 -  \frac{2\,m}{r} + \frac{m^2 \sin^2 \theta}{r^2},
\end{equation}
and
\begin{equation}\label{last3_label}
	\mu + \nu = 1 - \gamma.
\end{equation}
The $\gamma$ parameter represents the deviation from spherical symmetry, while the $\mu$ and $\nu$ parameters are associated with the presence of a scalar field in the metric. The total mass $M$ is given by $\gamma\,m$ and the scalar field is as follows
\begin{equation}\label{last4_label}
	\varphi(r) = \sqrt{\frac{1 - \gamma^2 - \nu}{2}} \; \ln \big( 1 - \frac{2 \, m}{r} \big).
\end{equation}
The metrics in Eq. \eqref{last1_label} 
reduce to the FJNW metric for $ \mu = 1 - \gamma $ and $ \nu = 0 $, and to the ZV metric for $ \mu = \gamma^2 - \gamma $ and $ \nu = 1 - \gamma^2. $\footnote{It has been recently pointed out in \cite{barrientos2024revisiting, barrientos2025rotating} that the exact solutions that were originally obtained in \cite{Azizallahi2023} can also be derived by applying Buchdahl's theorem \cite{buchdahl1956reciprocal, buchdahl1959reciprocal}.}

In the following, we derive a class of exact solutions to Einstein-Maxwell equations in the presence of a massless scalar field. To do this, we apply the Harrison transformations \cite{harrison1968new} to the metric \eqref{last1_label}. First, we rewrite the metric \eqref{last1_label} in the form of the Lewis-Weyl-Papapetrou (LWP) metric, which is given by
\begin{equation}\label{h1_label}
	ds^2=-F\,(dt-\omega\,d\phi)^2+F^{-1}\,\big[e^{2\,\lambda}\,(d\rho^2+dz^2)+\rho^2\,d\phi^2\big],
\end{equation}
where $ \rho $ and $ z $ can be expressed in terms of $ r $ and $ \theta $ as follows
\begin{equation}\label{h2_label}
	\rho=\sqrt{r\,(r-2\,m)}\,\sin\theta, \qquad z=(r-m)\,\cos\theta.
\end{equation}
Additionally, the vector potential is given by
\begin{equation}\label{h2b_label}
	A=A_t\,dt+A_\phi\,d\phi.
\end{equation}
Eri{\c{s}} and G{\"u}rses
\cite{erics1977stationary}
have shown that, by adding a scalar field to the stationary and axisymmetric Einstein-electrovacuum solutions, the degree of integrability remains the same as in the vacuum case, and only the $ \lambda $ function is modified in the LWP metric. Accordingly, the Ernst formulation remains fully applicable to Einstein's equations with a minimally coupled scalar field, exactly as in the vacuum scenario
\cite{erics1977stationary, astorino2013embedding, astorino2015stationary}.

By applying the coordinate transformation \eqref{h2_label} to the metric \eqref{h1_label} and comparing it with the metric \eqref{last1_label}, the functions $ F $, $ \omega $ and $ e^{2\,\lambda} $ are determined as follows
\begin{equation}\label{h3_label}
	\begin{aligned}
		&F(r) = (1 - \frac{2\,m}{r})^\gamma,\\
		&\omega=0,\\
		&e^{2\,\lambda}=\big[1+\frac{m^2\,\sin^2\theta}{r\,(r-2\,m)}\big]^{\nu-1}.
	\end{aligned}
\end{equation}
In 1968, Ernst reformulated the Einstein-Maxwell field equations into two compact equations, as presented in his papers \cite{ernst1968new, ernst1968new2}, in the following form
\begin{equation}\label{h4_label}
	\begin{aligned}
		&\big(\text{Re}(\varepsilon)+\Phi\,\Phi^\star\big)\,\nabla^2\,\varepsilon = \nabla\,\varepsilon\, .\, (\nabla\,\varepsilon+2\,\Phi^\star\,\nabla\,\Phi),\\
		&\big(\text{Re}(\varepsilon)+\Phi\,\Phi^\star\big)\,\nabla^2\,\Phi = \nabla\,\Phi\, .\, (\nabla\,\varepsilon+2\,\Phi^\star\,\nabla\,\Phi).
	\end{aligned}
\end{equation}
where $ Re(\varepsilon) $ represents the real part of the complex gravitational potential $ \varepsilon $ which is defined as follows
\begin{equation}\label{h5_label}
	\varepsilon:=F-\lvert\Phi\rvert^2 + i\,\chi,
\end{equation}
where $ F $ is the metric coefficient given by \eqref{h1_label}, and the functions $ \Phi $ and $ \chi $ are defined as follows
\begin{equation}\label{h6_label}
	\begin{aligned}
		&\Phi:=A_t+i\,\tilde{A}_\phi,\\
		&\hat{\phi}\times\nabla\,\chi:=-\rho^{-1}\,F^2\,\nabla\,\omega-2\,\hat{\phi}\times Im(\Phi^\star\,\nabla\,\Phi),\\
	\end{aligned}
\end{equation}
where $ Im(\Phi^\star\,\nabla\,\Phi) $ represents the imaginary part of the function $ \Phi^\star\,\nabla\,\Phi $ and $ \hat{\phi} $ is the unit vector in the $ \phi $ direction. Additionally, $ A_t $ represents the first component of $ A_\mu $ and $ \tilde{A}_\phi $ satisfies the following equation
\begin{equation}\label{h7_label}
	\hat{\phi}\times\nabla\,\tilde{A}_\phi:=\rho^{-1}\,F\,(\nabla\,A_\phi+\omega\,\nabla\,A_t),
\end{equation}
where $ A_\phi $ represents the fourth component of $ A_\mu $. In \cite{astorino2013embedding, astorino2015stationary}, Astorino demonstrated that, for the action in the presence of the scalar field, the scalar field decouples from Ernst's equations, and an additional equation related to the scalar field, ($ \Box \varphi (r) = 0 $), is added to Eqs. \eqref{h4_label}.

Additionally, the equations for the $ \lambda $ function, according to the Ernst potentials, are given by
\begin{equation}\label{h7a_label}
	\begin{aligned}
		\partial_\rho\,\lambda=&\frac{\rho}{4\,(Re(\varepsilon)+\Phi\,\Phi^*)^2}\,\big[(\partial_\rho\,\varepsilon+2\,\Phi^*\,\partial_\rho\,\Phi)\,(\partial_\rho\,\varepsilon^*+2\,\Phi\,\partial_\rho\,\Phi^*)-(\partial_z\,\varepsilon+2\,\Phi^*\,\partial_z\,\Phi)\,(\partial_z\,\varepsilon^*\\
		&+2\,\Phi\,\partial_z\,\Phi^*)\big]-\frac{\rho}{Re(\varepsilon)+\Phi\,\Phi^*}\,\big[\partial_\rho\,\Phi\,\partial_\rho\,\Phi^*-\partial_z\,\Phi\,\partial_z\,\Phi^*\big]+\frac{\rho}{2}\,\big[(\partial_\rho\,\varphi)^2-(\partial_z\,\varphi)^2\big],
	\end{aligned}
\end{equation}
\begin{equation}\label{h7b_label}
	\begin{aligned}
		\partial_z\,\lambda=&\frac{\rho}{4\,(Re(\varepsilon)+\Phi\,\Phi^*)^2}\,\big[(\partial_\rho\,\varepsilon+2\,\Phi^*\,\partial_\rho\,\Phi)\,(\partial_z\,\varepsilon^*+2\,\Phi\,\partial_z\,\Phi^*)+(\partial_z\,\varepsilon+2\,\Phi^*\,\partial_z\,\Phi)\,(\partial_\rho\,\varepsilon^*\\
		&+2\,\Phi\,\partial_\rho\,\Phi^*)\big]-\frac{\rho}{Re(\varepsilon)+\Phi\,\Phi^*}\,\big[\partial_\rho\,\Phi\,\partial_z\,\Phi^*-\partial_z\,\Phi\,\partial_\rho\,\Phi^*\big]+\rho\,\big[\partial_\rho\,\varphi\,\partial_z\,\varphi\big],
	\end{aligned}
\end{equation}
We now introduce the Harrison transformations, defined as follows
\begin{equation}\label{h8_label}
	\varepsilon^\prime = \frac{\varepsilon}{1-2\,\alpha^\star\,\Phi-\lvert\alpha\rvert^2\,\varepsilon}, \qquad \Phi^\prime=\frac{\alpha\,\varepsilon+\Phi}{1-2\,\alpha^\star\,\Phi-\lvert\alpha\rvert^2\,\varepsilon},
\end{equation}
where $ \alpha $ is a complex number. Considering the metric \eqref{last1_label} and using Eqs. \eqref{h3_label}, \eqref{h5_label}, and \eqref{h6_label}, we obtain
\begin{equation}\label{h9_label}
	\varepsilon=(1 - \frac{2\,m}{r})^\gamma, \qquad \Phi=0, \qquad \chi=0.
\end{equation}
By substituting Eqs. \eqref{h9_label} into Eqs. \eqref{h8_label}, we obtain
\begin{equation}\label{h10_label}
	\begin{aligned}
		&\varepsilon^\prime=\frac{(r-2\,m)^\gamma}{r^\gamma-\lvert\alpha\rvert^2\,(r-2\,m)^\gamma},\\
		&\Phi^\prime=\frac{\alpha\,(r-2\,m)^\gamma}{r^\gamma-\lvert\alpha\rvert^2\,(r-2\,m)^\gamma}.
	\end{aligned}
\end{equation}
Using Eq. \eqref{h5_label}, the functions $ F^\prime $ and $ \chi^\prime $ are determined by the following equations
\begin{equation}\label{h11_label}
	F^\prime=\varepsilon^\prime+\lvert\Phi^\prime\rvert^2=\frac{\big[r\,(r-2\,m)\big]^\gamma }{\big[r^\gamma-(r-2\,m)^{\gamma}\;\lvert\alpha\rvert^2\big]^2}, \qquad \chi^\prime=0.
\end{equation}
Now, by defining the parameter $\alpha$ as follows
\begin{equation}\label{h12_label}
	\alpha=\alpha_R+i\,\alpha_I,
\end{equation}
and using Eqs. \eqref{h6_label} and \eqref{h10_label}, we can define $ A_t^\prime $ and $ \tilde{A}_\phi^\prime $ as follows
\begin{equation}\label{h13_label}
	A_t^\prime=\frac{\alpha_R\,(r-2\,m)^\gamma}{r^\gamma-\lvert\alpha\rvert^2\,(r-2\,m)^\gamma}, \qquad \tilde{A}_\phi^\prime=\frac{\alpha_I\,(r-2\,m)^\gamma}{r^\gamma-\lvert\alpha\rvert^2\,(r-2\,m)^\gamma}.
\end{equation}
Now, to find $ A_\phi^\prime $, we use Eq. \eqref{h7_label}. In using Eqs. \eqref{h6_label} and \eqref{h7_label}, it should be noted that all the gradients in these equations are in the cylindrical coordinate system, and to use them in the spherical coordinate system, one can use the below transformation
\begin{equation}\label{h14_label}
	\nabla\,h(r,\,\theta)=\frac{1}{\sqrt{(r-m)^2-m^2\,\cos^2\theta}}\;\big[\frac{\partial\,h(r,\,\theta)}{\partial\,r}\,\sqrt{r\,(r-2\,m)}\,\hat{r}+\frac{\partial\,h(r,\,\theta)}{\partial\,\theta}\,\hat{\theta}\,\big],
\end{equation}
where $h(r,\,\theta)$ is an arbitrary differentiable function of $r$ and $\theta$. Now, by using Eqs. \eqref{h2_label}, \eqref{h7_label}, \eqref{h11_label}, \eqref{h13_label}, and \eqref{h14_label}, we obtain $ A_\phi^\prime $ as follows
\begin{equation}\label{h15_label}
	A_\phi^\prime=2\,\gamma\,m\,\alpha_I\,\cos\theta.
\end{equation}
The final form of the vector potential is given by
\begin{equation}\label{h16_label}
	A^\prime=\frac{\alpha_R\,(r-2\,m)^\gamma}{r^\gamma-\lvert\alpha\rvert^2\,(r-2\,m)^\gamma}\,dt+2\,\gamma\,m\,\alpha_I\,\cos\theta\,d\phi.
\end{equation}
Therefore, according to Eqs. \eqref{h1_label}, \eqref{h2_label}, \eqref{h3_label}, and \eqref{h11_label}, the metric in the presence of the electromagnetic field can be written as follows (note that the Harrison transformations do not affect the $ \lambda $ function, and the reason for this will be explained shortly):
\begin{equation}\label{h17_label}
	\begin{aligned}
		d s^\prime{^2} = & - \frac{\big[r\,(r-2\,m)\big]^\gamma }{\big[r^\gamma-(r-2\,m)^{\gamma}\;\lvert\alpha\rvert^2\big]^2} \, d t^2 + 
		\big[r^\gamma-(r-2\,m)^{\gamma}\;\lvert\alpha\rvert^2\big]^2 \, \big[r\,(r-2\,m)\big]^{\mu-1}\\
		&\times \big[r\,(r-2\,m)+m^2\,\sin^2\theta\big]^\nu \, \big\{ d r^2 + [r\,(r-2\,m)]\, d\theta^2\big\}+ 
		\big[r^\gamma-(r-2\,m)^{\gamma}\;\lvert\alpha\rvert^2\big]^2 \\
		&\times \big[r\,(r-2\,m)\big]^{1-\gamma} \, \sin^2 \theta \, d \phi^2.
	\end{aligned}
\end{equation}
We now apply the following coordinate transformation
\begin{equation}\label{h18_label}
	\begin{aligned}
		&\tilde{r}=r(1-\lvert\alpha\rvert^2)+2\,m\,\lvert\alpha\rvert^2, \qquad \tilde{t}=\frac{t}{\big(1-\lvert\alpha\rvert^2\big)},\\
		&\tilde{m}=m\,(1+\lvert\alpha\rvert^2), \qquad q_e=-2\,m\,\alpha_R,\qquad q_m=2\,m\,\alpha_I.
	\end{aligned}
\end{equation}
Here, $ q_e $ and $ q_m $ are proportional to the electric and magnetic charges, respectively. For the sake of simplicity in calculations, we also define a parameter $ p $ as follows
\begin{equation}\label{h19_label} p:=\sqrt{1-\frac{q_e^2+q_m^2}{\tilde{m}^2}}.
\end{equation}
Using the transformations and definitions in \eqref{h18_label} and \eqref{h19_label}, we express the coefficient of $dt^2$ as $F^\prime=\mathcal{F}^\gamma/\mathcal{R}^2$, and remove the tilde from the parameters ($\tilde{r}\rightarrow r$, $\tilde{t}\rightarrow t$ and $\tilde{m}\rightarrow m$). Finally, the metric \eqref{h17_label} can be written as follows
\begin{equation}\label{1_label}
	d s^2 = - \, \frac{\mathcal{F}^\gamma}{\mathcal{R}^2} \, d t^2 + \mathcal{R}^2 \, \mathcal{F}^\mu \, \mathcal{K}^\nu \, \big[ \frac{1}{\mathcal{F}} \, d r^2 + r^2 \, d \theta^2 \big] + \mathcal{R}^2 \, \mathcal{F}^{1 - \gamma} \, r^2 \, \sin^2 \theta \, d \phi^2,
\end{equation}
where 
\begin{equation}\label{3_label}
	\begin{aligned}
		&\mu+\nu=1-\gamma,\\
		& \mathcal{F}(r) = 1 - \frac{2\,m}{r} + \frac{q_e^2+q_m^2}{r^2},\\ 
		& \mathcal{K}(r,\theta) = 1 - \frac{2\,m}{r} + \frac{q_e^2+q_m^2}{r^2} + \frac{(m^2-q_e^2-q_m^2) \, \sin^2\theta}{r^2},\\
		&\mathcal{R}(r)=\frac{1+p}{2\,p}\,\Big[\big(1-\frac{m\,(1-p)}{r}\big)^\gamma-\frac{1-p}{1+p}\;\big(1-\frac{m\,(1+p)}{r}\big)^\gamma\Big].
	\end{aligned}
\end{equation}
The asymptotic behavior of the metrics implies that the physical mass and the squares of the physical electric and magnetic charges are given by $ M=\gamma \, m $, $ Q_e^2=\gamma \, q_e^2 $ and $ Q_m^2=\gamma \, q_m^2 $, respectively. Additionally, the vector potential $ A^\prime $, can be written in the following form
\begin{equation}\label{h50_label}
	A^\prime=A_0(r)\,dt+q_m\,\gamma\,\cos\theta\,d\phi,
\end{equation}
where
\begin{equation}\label{h51_label}
	\begin{aligned}
		A_0(r)&=\frac{q_e}{(1+p)\,m\,\mathcal{R}}\;\big(1-\frac{(1+p)\,m}{r}\big)^\gamma,\\
		&=\gamma\,q_e\,\int\,\frac{\mathcal{F}^{\gamma-1}}{\mathcal{R}^2\,r^2}.
	\end{aligned}
\end{equation}
Based on Eq. \eqref{101b_label} and metric \eqref{1_label}, we now derive the scalar field as follows
\begin{equation}\label{103_label}
	\varphi(r) = \sqrt{\frac{1 - \gamma^2 - \nu}{2}} \; \ln \big(\frac{r - m - \sqrt{m^2 - q_e^2- q_m^2}}{r - m +\sqrt{m^2 - q_e^2- q_m^2}} \big).
\end{equation}
By using the following coordinate transformations:
\begin{equation}\label{103a_label}
	\begin{aligned}
		&\theta=\cos^{-1}\big(\frac{R_+-R_-}{2\,m}\big),\qquad r=m+\frac{1}{2}\,(R_++R_-),\\ &R_+=\sqrt{\rho^2+(z+m)^2},\qquad R_-=\sqrt{\rho^2+(z-m)^2},
	\end{aligned}
\end{equation}
Using Eqs. \eqref{last4_label}, \eqref{h9_label}, \eqref{h10_label}, \eqref{h18_label}, and \eqref{103_label} in Eqs. \eqref{h7a_label} and \eqref{h7b_label}, we conclude that the Harrison transformations do not affect the $\lambda$ function, and thus $ \lambda=\lambda^\prime $.

According to Eqs. \eqref{last3_label} and \eqref{103_label}, we obtain
\begin{equation}\label{201_label}
	\nu \leq 1 - \gamma^2, \qquad \mu \geq \gamma^2 - \gamma\geq-\frac{1}{4}.
\end{equation}
The metrics presented in Eq. \eqref{1_label} constitute a class of Einstein-Maxwell solutions in the presence of a scalar field, which reduce to interesting known solutions in certain limits. If we set $ \mu $ and $ \nu $ as below
\begin{equation}\label{b_label}
	\mu = \gamma^2 - \gamma, \qquad \nu = 1 - \gamma^2,
\end{equation}
metric \eqref{1_label} becomes the charged  ZV metric. Now by choosing $ \nu = 0 $ ($\mu=1-\gamma$), metric \eqref{1_label} represents the charged FJNW metric. When the electric and magnetic charges vanish ($q_e = q_m = 0$), metric \eqref{1_label} reduces to metric \eqref{last1_label}, representing a spacetime with a scalar field, which is therefore not an electrovacuum solution of Einstein's equations.

By choosing $ \gamma = 1 $, $ \mathcal{R}(r) $ in Eq. \eqref{3_label} becomes equal to one and the metric in Eq. \eqref{1_label} simplifies to
\begin{equation}\label{710c_label}
	\begin{aligned}
		ds^2=&-(1-\frac{2\,m}{r}+\frac{q^2}{r^2})\,dt^2+\big[1+\frac{(m^2-q^2)\,\sin^2\theta}{r^2-2\,m\,r+q^2}\big]^\nu\;\big[\frac{dr^2}{1-2\,m/r+q^2/r^2}+r^2\,d\theta^2\big]\\
		&+r^2\,\sin^2\theta\,d\phi^2.
	\end{aligned}
\end{equation}
Additionally, in this case, by setting $ q_m=0 $, using Eqs. \eqref{h6_label}, \eqref{h50_label}, and \eqref{h51_label}, and omitting a constant term in the vector potential $ A^\prime $, we obtain
\begin{equation}\label{710cd_label}
	A^\prime=\frac{q}{r}\,dt,\quad\Rightarrow\quad\Phi^\prime=\frac{q}{r}.
\end{equation}
When the parameter $\nu$ vanishes, metric \eqref{710c_label} reduces to the Reissner-Nordstr\"{o}m metric.

In the following, we will examine the properties of metric \eqref{1_label} and its effects on the surrounding environment.
\subsection{Singularities of the Metrics}\label{2a}
First, we examine the metric singularities by calculating the Ricci scalar. The exact form of the Ricci scalar for metric \eqref{1_label} is derived as follows
\begin{equation}\label{5_label}
	R(r,\,\theta)=-\frac{8\,m^2\,p^4\,[\mu^2+(2\,\mu+\nu)\,(\nu-1)]}{r^{2\,\gamma}\,Y(r)^2\,[r^2-2\,m\,r+q_e^2+q_m^2]^{\mu+1}\;[r^2-2\,m\,r+m^2\,\sin^2\theta+(q_e^2+q_m^2)\,\cos^2\theta]^\nu},
\end{equation}
where $ Y(r,\,\theta) $ is a regular function as follows
\begin{equation}\label{5a_label}
	Y(r)=(1+p)\,[1+\frac{m\,(p-1)}{r}]^\gamma-(1-p)\,[1-\frac{m\,(p+1)}{r}]^\gamma.
\end{equation}

Based on Eq. \eqref{5_label}, the singularity points can be expressed as follows
\begin{equation}\label{5b_label}
	r_{singularity}=\begin{cases}
		0,\qquad \qquad \qquad \qquad \quad \qquad\qquad\qquad  \;\text{for}\; \gamma>0,\\m\pm \sqrt{m^2-(q_e^2+q_m^2)},\qquad \qquad \qquad \quad \;\text{for}\; \mu>-1,\\m\pm\cos\theta\, \sqrt{m^2-(q_e^2+q_m^2)},\qquad\qquad\quad \text{for}\; \nu>0.
	\end{cases}
\end{equation}
According to Eq. \eqref{201_label}, $ \mu $ can never be smaller than or equal to -1, so a singularity always occurs at $m\pm \sqrt{m^2-(q_e^2+q_m^2)}$. Additionally, if  $\nu \le 0$, no singularity is present at $m\pm\cos\theta\, \sqrt{m^2-(q_e^2+q_m^2)}$.

\begin{figure}[ht]
\centering
\includegraphics[width=0.6\linewidth]{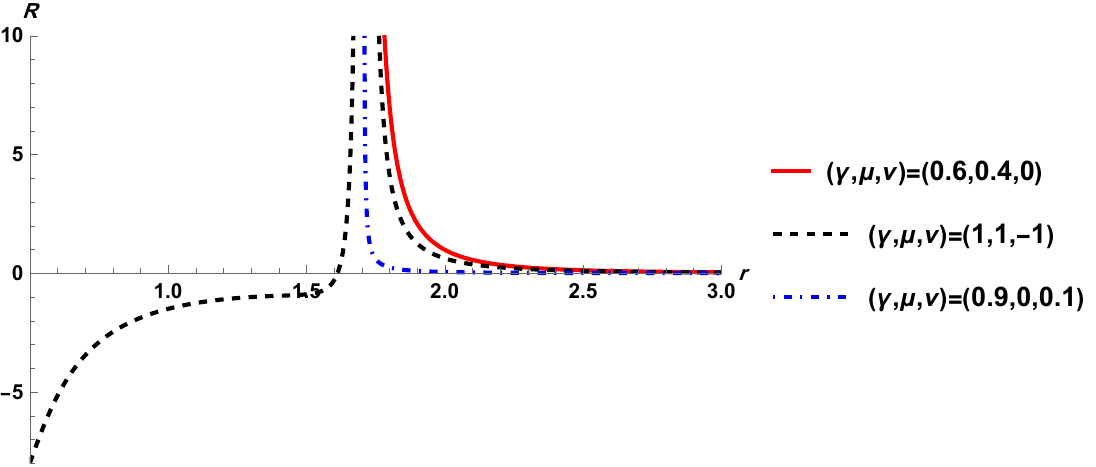}
\caption[Plot of the Ricci scalar]{\small Plot of the Ricci scalar in Eq.~\eqref{5_label} as a function of $r$, for $m = 1$, $q_e = q_m = 1/2$, and $\theta = \pi/6$. Singularities occur at $m + \sqrt{m^2 - (q_e^2 + q_m^2)}$, and the Ricci scalar approaches zero at large $r$.}
\label{riccich}
\end{figure}

Therefore, in charged metrics \eqref{1_label}, all singularities occur at $ r\le m+\sqrt{m^2-(q_e^2+q_m^2)} $. As we approach the gravitational source from the outer region $ r>m+\sqrt{m^2-(q_e^2+q_m^2)} $, we encounter a naked singularity at $ r=m+\sqrt{m^2-(q_e^2+q_m^2)} $, where the volume element vanishes. In Fig. \ref{riccich}, the Ricci scalar corresponding to metric \eqref{1_label} is plotted as a function of $ r$ for different values of $ \gamma $, $ \mu $ and $ \nu $.
\subsection{Geodesics and Effective Potential}\label{4}
The study of geodesics and the motion of neutral particles in a spacetime provides valuable insights into the characteristics of that spacetime. In this section, we first derive the geodesic equations for the metrics in \eqref{1_label}, and then we proceed to obtain the effective potential for neutral particles in the background of these metrics.
\subsubsection{Geodesics}\label{4a}
In this section, we will determine the permissible range of the $ \gamma $ parameter by deriving the geodesic equations for the metric in \eqref{1_label}.

The constraints and equations governing the geodesic curves are as follows
\begin{equation}\label{g1_label}
	\ddot{x}^\alpha+\Gamma_{\beta\,\sigma}^\alpha\,\dot{x}^\beta\,\dot{x}^\sigma=0,
\end{equation}
\begin{equation}\label{kh2_label}
	g_{\alpha\,\beta}\,\dot{x}^\alpha\,\dot{x}^\beta=-\eta,\qquad
	\eta=-1,\,0,\,1,
\end{equation} 
Here, the dot represents the derivative with respect to the affine parameter $ \tau $, $ x^\alpha $ denotes the spacetime coordinates along the geodesics, and $ \eta $ is defined as
\begin{equation}\label{kh3_label}
	\eta=\begin{cases}
		-1,\qquad\,\,\,  \text{for}\; \text{spacelike curves},\\0,\quad\qquad  \text{for}\; \text{null curves},\\1,\quad\qquad  \text{for}\; \text{timelike curves}.
	\end{cases}
\end{equation}
Using Eq. \eqref{g1_label}, the geodesic equations for the metric in \eqref{1_label} are derived as follows
\begin{equation}\label{g3_label}
	\ddot{t}+\big(\frac{\gamma\,\partial_r\,\mathcal{F}}{\mathcal{F}}-\frac{2\,\partial_r\,\mathcal{R}}{\mathcal{R}}\big)\,\dot{t}\,\dot{r}=0,
\end{equation}
\begin{equation}\label{g4_label}
	\begin{aligned}
		&\ddot{r}+\frac{\mathcal{F}^{\gamma-\mu}\,\mathcal{K}^{-\nu}\,(\gamma\,\mathcal{R}\,\partial_r\,\mathcal{F}-2\,\mathcal{F}\,\partial_r\,\mathcal{R})}{2\,\mathcal{R}^5}\,\dot{t}^2+\frac{1}{2}\,\big(\frac{(\mu-1)\,\partial_r\,\mathcal{F}}{\mathcal{F}}+\frac{2\,\partial_r\,\mathcal{R}}{\mathcal{R}}+\frac{\nu\,\partial_r\,\mathcal{K}}{\mathcal{K}}\big)\,\dot{r}^2+\frac{\nu\,\partial_\theta\,\mathcal{K}}{\mathcal{K}}\,\dot{r}\,\dot{\theta}\\
		&-\frac{1}{2}\,r\,\big[\mathcal{F}\,\big(2+\frac{2\,r\,\partial_r\,\mathcal{R}}{\mathcal{R}}+\frac{r\,\nu\,\partial_r\,\mathcal{K}}{\mathcal{K}}\big)+r\,\mu\,\partial_r\,\mathcal{F}\big]\dot{\theta}^2+\big(\frac{\mathcal{F}}{\mathcal{K}}\big)^\nu\,\frac{r\,(\gamma-1)\,\mathcal{R}\,\partial_r\,\mathcal{F}-2\,\mathcal{F}\,(\mathcal{R}+r\,\partial_r\,\mathcal{R})}{2\,\mathcal{R}}\\
		&\times r\,\sin^2\theta\,\dot{\phi}^2=0,
	\end{aligned}
\end{equation}
\begin{equation}\label{g5_label}
	\ddot{\theta}-\frac{\nu\,\partial_\theta\,\mathcal{K}}{2\,r^2\,\mathcal{F}\,\mathcal{K}}\,\dot{r}^2+\big[\frac{2}{r}+\frac{\mu\,\partial_r\,\mathcal{F}}{\mathcal{F}}+\frac{\nu\,\partial_r\,\mathcal{K}}{\mathcal{K}}+\frac{2\,\partial_r\,\mathcal{R}}{\mathcal{R}}\big]\,\dot{r}\,\dot{\theta}+\frac{\nu\,\partial_\theta\,\mathcal{K}}{2\,\mathcal{K}}\,\dot{\theta}^2-\frac{1}{2}\,\big(\frac{\mathcal{F}}{\mathcal{K}}\big)^\nu\,\sin(2\,\theta)\,\dot{\phi}^2=0,
\end{equation}
\begin{equation}\label{g6_label}
	\ddot{\phi}+\big[\frac{2}{r}-\frac{(\gamma-1)\,\partial_r\,\mathcal{F}}{\mathcal{F}}+\frac{2\,\partial_r\,\mathcal{R}}{\mathcal{R}}\big]\,\dot{r}\,\dot{\phi}+2\,\cot\theta\,\dot{\theta}\,\dot{\phi}=0.
\end{equation}
We assume that a geodesic is described by the following equation
\begin{equation}\label{g7a_label}
	\sigma(\tau)=\big(t(\tau),\,r(\tau),\,\theta(\tau),\,\phi(\tau)\big).
\end{equation}
Without losing generality, we can express Eq. \eqref{g7a_label} as follows
\begin{equation}\label{g7b_label}
	\sigma(\tau_0)=\big(t(\tau_0),\,r(\tau_0),\,\frac{\pi}{2},\,\phi(\tau_0)\big).
\end{equation}
Additionally, from Eq. \eqref{3_label}, we know that $ \partial_\theta\mathcal{K} $  is proportional to $ \sin(2\theta) $. By choosing $ \theta(\tau_0)=\pi/2 $, the solution to Eq. \eqref{g5_label} is $ \dot{\theta}=0 $. Using this condition and integrating Eqs. \eqref{g3_label} and \eqref{g6_label}, we obtain the following two equations, respectively
\begin{equation}\label{g8_label}
	\dot{t}=c_1\,\mathcal{F}^{-\gamma}\,\mathcal{R}^2,
\end{equation}
\begin{equation}\label{g9_label}
	\dot{\phi}=c_2\,\frac{\mathcal{F}^{\gamma-1}}{r^2\,\mathcal{R}^2},
\end{equation}
where $ c_1 $ and $ c_2 $ are constants. By considering the condition $ \theta = \pi/2 $, $ p $ is equal to $ \sqrt{1 - q_e^2 / m^2} $, and as a result, we have $ \lvert q_e \rvert \le m $. Also, by using Eqs. \eqref{g8_label} and \eqref{g9_label}, the square of the angular velocity is determined by the following equation
\begin{equation}\label{23_label}
	\begin{aligned}
		\omega^2 = &\frac{\dot{\phi}^2}{\dot{t}^2}= \frac{c_2^2}{c_1^2\,\mathcal{R}^8 \, r^4} \big( 1 - \frac{2 \, m}{r} + \frac{q_e^2}{r^2} \big)^{4 \, \gamma - 2}\\
		= & \frac{1}{\mathcal{R}^8 \, r^{8 \, \gamma}} \big( (r - r_+) \, (r - r_-) \big)^{4 \, \gamma - 2}.
	\end{aligned}
\end{equation}
where $ r_+ $ and $ r_- $ are as follows
\begin{equation}\label{24_label}
	r_+ = m \, (1 + p), \qquad r_- = m \, (1 - p).
\end{equation}
Eq. \eqref{23_label} shows that for $ \gamma < 1/2 $, $ \omega^2 $ tends to infinity as $ r \rightarrow r_\pm $. Therefore, for $ \gamma \ge 1/2 $ ($ \mu + \nu \le 1/2 $), $ \omega $ remains finite for all values of the parameters. In the following, we will investigate the motion of neutral particles in the background of charged metrics \eqref{1_label}.
\subsubsection{Effective Potential of Neutral Particles}\label{4b}
In this section, we examine the behavior of the effective potential for neutral particles and explore the impact of charge and the scalar field on this potential.

\begin{figure}[ht]
\centering
\subfloat[]{\includegraphics[width=0.44\linewidth]{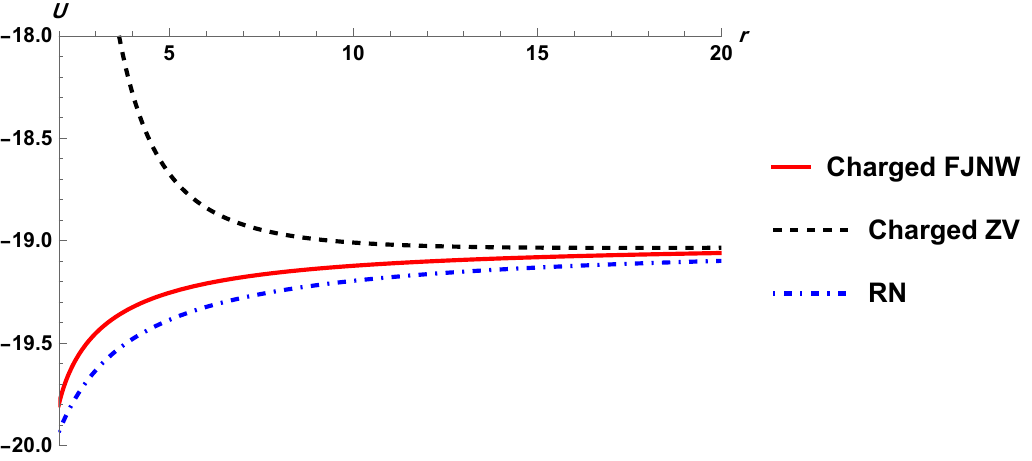}}
\qquad
\subfloat[]{\includegraphics[width=0.44\linewidth]{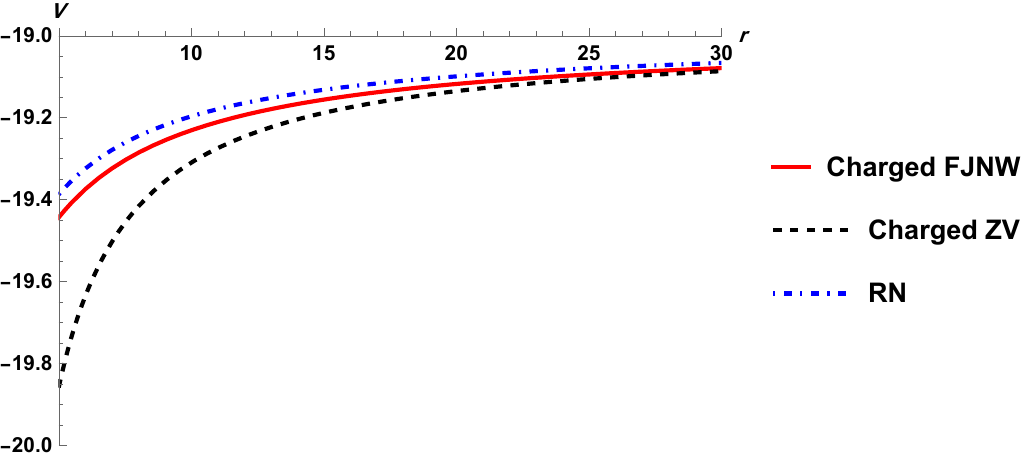}}
\qquad
\subfloat[]{\includegraphics[width=0.44\linewidth]{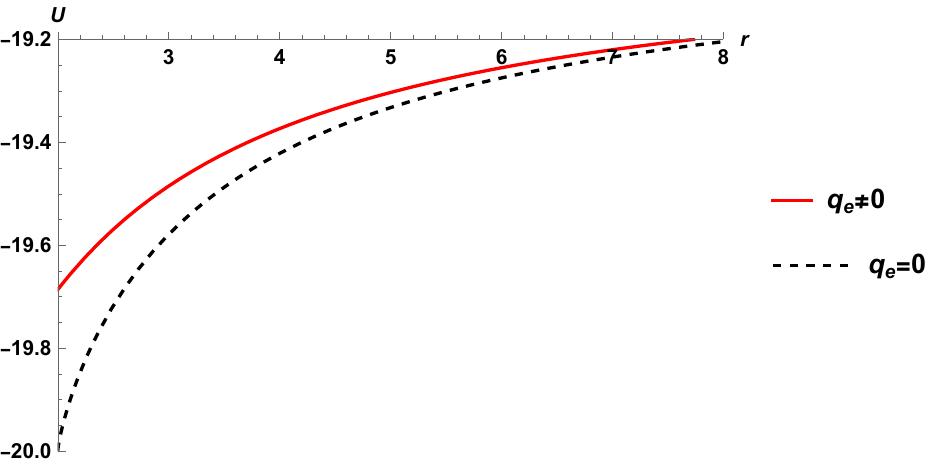}}
\caption[]{\small The plot of the effective potential in Eq.~\protect\eqref{kh10_label} as a function of $r$ is shown for $m = 1$, $\eta = 1$, $E^2/m_0^2 = 20$, and $L^2/m_0^2 = 0.1$. Plot (a) depicts graphs for the case where $\gamma \le 1$. For $q_e = 0.5$, $\gamma = 0.6$, $\mu = 0.4$, and $\nu = 0$, the charged FJNW metric is represented by a solid red line. For $\gamma = 0.6$, $\mu = -0.24$, and $\nu = 0.64$, the charged ZV metric is shown by a black dashed line. For $\gamma = 1$, $\mu = 0$, and $\nu = 0$, the RN metric is shown as a blue dot-dashed line. Plot (b) illustrates graphs for the case where $\gamma \ge 1$. For $q_e = 0.5$, $\gamma = 1.2$, $\mu = -0.2$, and $\nu = 0$, the charged FJNW metric is shown by a solid red line. For $\gamma = 1.2$, $\mu = 0.24$, and $\nu = -0.44$, the charged ZV metric is shown by a black dashed line. For $\gamma = 1$, $\mu = 0$, and $\nu = 0$, the RN metric is represented by a blue dot-dashed line. Plot (c) shows the case where $\gamma = 0.8$, $\mu = 0.2$, and $\nu = 0$. For $q_e = 0.99$, the charged FJNW metric is shown by a solid red line, and for $q_e = 0$, the FJNW metric is represented by a black dashed line.}
\label{figveff}
\end{figure}

According to Eq. \eqref{kh2_label}, for the metric in Eq. \eqref{1_label}, we obtain:
\begin{equation}\label{kh4_label}
	-\frac{\mathcal{F}^\gamma}{\mathcal{R}^2} \, \dot{t}^2 + \mathcal{R}^2 \, \mathcal{F}^{\mu-1} \, \mathcal{K}^\nu \,  \dot{r}^2+\mathcal{R}^2 \, \mathcal{F}^\mu \, \mathcal{K}^\nu \,r^2\,  \dot{\theta}^2 + \mathcal{R}^2 \, \mathcal{F}^{1 - \gamma} \, r^2 \,\sin^2\theta\, \dot{\phi}^2=-\eta.
\end{equation}
The following Lagrangian describes the motion of a neutral particle with mass $m_0$ in the background of the metric given by Eq. \eqref{1_label}:
\begin{equation}\label{kh5_label}
	\begin{aligned}
		\mathcal{L}&=\frac{1}{2}\,g_{\alpha\,\beta}\,\dot{x}^\alpha\,\dot{x}^\beta\\	
		&=- \, \frac{\mathcal{F}^\gamma}{2\,\mathcal{R}^2} \, \dot{t}^2 +\frac{1}{2}\, \mathcal{R}^2 \, \mathcal{F}^{\mu-1} \,\mathcal{K}^\nu \,  \dot{r}^2+\frac{1}{2}\, \mathcal{R}^2 \, \mathcal{F}^\mu \, \mathcal{K}^\nu \,r^2\,  \dot{\theta}^2 +\frac{1}{2}\, \mathcal{R}^2 \, \mathcal{F}^{1 - \gamma} \, r^2\,\sin^2\theta\, \dot{\phi}^2.
	\end{aligned}
\end{equation}
According to the Lagrangian \eqref{kh5_label}, we have the following equations for the momentum of particles:
\begin{equation}\label{kh6_label}
	p_t=\frac{\partial\,\mathcal{L}}{\partial\,\dot{t}}=-\frac{\mathcal{F}^\gamma\,\dot{t}}{\mathcal{R}^2}=-\frac{E}{m_0},\quad\Rightarrow\quad\dot{t}=\frac{E\,\mathcal{R}^2}{m_0\,\mathcal{F}^\gamma},
\end{equation}
\begin{equation}\label{kh7_label}
	p_\phi=\frac{\partial\,\mathcal{L}}{\partial\,\dot{\phi}}=\mathcal{R}^2 \, \mathcal{F}^{1 - \gamma} \, r^2 \, \sin^2\theta\,\dot{\phi}=\frac{L}{m_0},\quad\Rightarrow\quad\dot{\phi}=\frac{L}{m_0\,\mathcal{R}^2 \, \mathcal{F}^{1 - \gamma} \, r^2\,\sin^2\theta},
\end{equation}
Where $ E $ and $ L $ are the energy and angular momentum of the particles, respectively. In this part, as in the previous one, we limit our study to equatorial geodesics, i.e., $ \theta = \pi/2 $. Then, by using Eqs. \eqref{kh6_label} and \eqref{kh7_label} in Eq. \eqref{kh4_label}, we have
\begin{equation}\label{kh8_label}
	\dot{r}^2 + \frac{L^2}{m_0^2\,\mathcal{R}^4 \, \mathcal{F}^{\mu - \gamma}\, \mathcal{K}^\nu \, r^2}+\frac{\eta}{\mathcal{R}^2 \, \mathcal{F}^{\mu-1} \, \mathcal{K}^\nu}-\frac{E^2}{m_0^2\,\mathcal{F}^{-\nu}\,\mathcal{K}^\nu}=0.
\end{equation}
By rewriting Eq. \eqref{kh8_label} as below
\begin{equation}\label{kh9_label}
	\dot{r}^2 +U(r)=0.
\end{equation}
The effective potential that neutral particles are affected by is described as follows
\begin{equation}\label{kh10_label}
	U(r)=\frac{L^2}{m_0^2\,\mathcal{R}^4 \, \mathcal{F}^{\mu - \gamma}\, \mathcal{K}^\nu \, r^2}+\frac{\eta}{\mathcal{R}^2 \, \mathcal{F}^{\mu-1} \, \mathcal{K}^\nu}-\frac{E^2}{m_0^2\,\mathcal{F}^{-\nu}\,\mathcal{K}^\nu}.
\end{equation}
According to Eq. \eqref{kh9_label}, one-dimensional motion is possible ($ \dot{r} \neq 0 $) only in regions where $ U < 0 $, which occurs when $ E $ in Eq. \eqref{kh10_label} is large enough. To investigate the motion of neutral particles in metrics \eqref{1_label}, we need to study the effective potential \eqref{kh10_label}. Next, we will examine the effects of the charge and scalar field on the effective potential of neutral particles. For this purpose, we have presented diagrams of the effective potential as a function of the radius, both in the presence and absence of the scalar field, and in the presence and absence of the electric charge, as shown in Fig. \ref{figveff}.

According to Fig. \ref{figveff}, in the case of $ \gamma \le 1 $, the effective potential of neutral particles in the background of the charged ZV metric at distances close to the source is greater than the effective potential in the presence of RN and FJNW metrics. As expected, at far distances from the source, the effective potential reaches a constant value for all three metrics. In the case of $ \gamma \ge 1 $, unlike the previous case, the effective potential in the background of the charged ZV metric at distances close to the source is smaller than the effective potential in the presence of RN and FJNW metrics. Also, the presence of electric charge increases the effective potential near the gravitational source.

\begin{figure}[ht]
\centering
\subfloat{\includegraphics[width=0.7\linewidth]{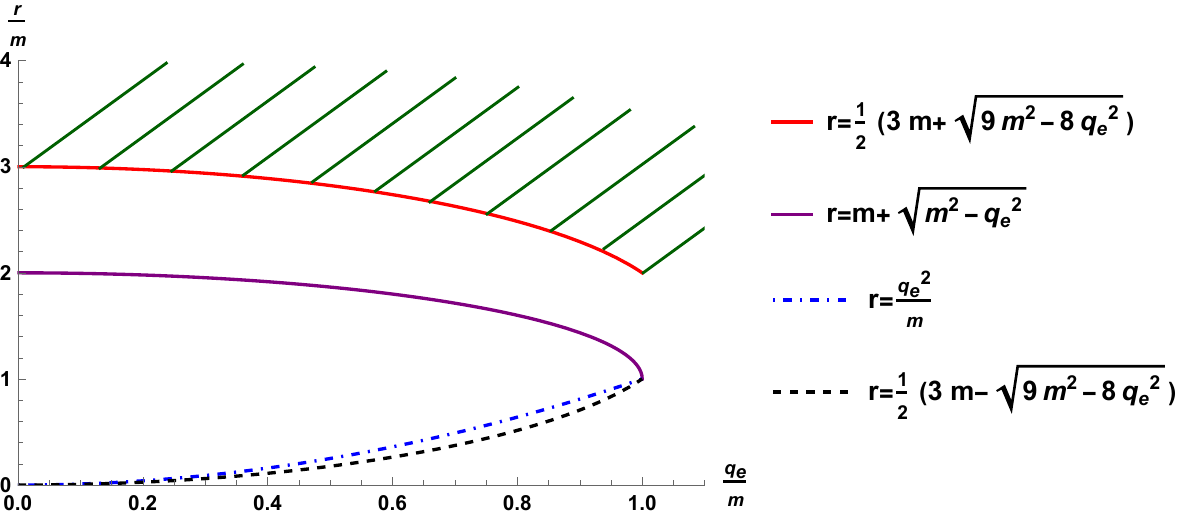}}
\caption[]{\small Graphs of $r/m$ versus $q_e/m$ are depicted according to Eqs.~\protect\eqref{kh16_label}. According to these diagrams, the minimum distance allowed from the source for the circular motion of neutral particles must be lower than the black dashed diagram, higher than the solid red diagram, and also higher than the solid violet and the blue dot-dashed line. Therefore, the green dashed areas in the figure (above the red graph) meet the conditions.}
\label{radi}
\end{figure}

Now, we limit our study to the circular motion of particles in the timelike curves where $r$ is constant ($ \dot{r} = 0 $) and $ \eta = 1 $. For this purpose, to simplify the calculations, we multiply both sides of Eq. \eqref{kh8_label} by $ \mathcal{F}^{-\nu} , \mathcal{K}^\nu $ and define a new effective potential at a fixed radius as follows
\begin{equation}\label{kh11_label}
	U_r(r)=\frac{L^2}{m_0^2\,\mathcal{R}^4\,\mathcal{F}^{1-2\,\gamma}\,r^2}+\frac{\eta}{\mathcal{R}^2\,\mathcal{F}^{-\gamma}}-\frac{E^2}{m_0^2}.
\end{equation}
Now we are going to obtain the angular momentum and energy of the particles. For circular orbits, we set $ \partial_r , U_r(r) = 0 $ and we have
\begin{equation}\label{kh12_label}
	\frac{L^2}{m_0^2}=\frac{\eta\,r^3\,\mathcal{R}^2\,\mathcal{F}^{1-\gamma}\,(\gamma\,\mathcal{R}\,\partial_r\,\mathcal{F}-2\,\mathcal{F}\,\partial_r\,\mathcal{R})}{(1-2\,\gamma)\,r\,\mathcal{R}\,\partial_r\,\mathcal{F}+2\,\mathcal{F}\,(\mathcal{R}+2\,r\,\partial_r\,\mathcal{R})}.
\end{equation}
By using Eqs. \eqref{kh9_label}, \eqref{kh11_label}, and \eqref{kh12_label}, we have:
\begin{equation}\label{kh13_label}
	\frac{E^2}{m_0^2}=\frac{\eta\,\mathcal{F}^\gamma\,\big[2\,r\,\mathcal{F}\,\partial_r\,\mathcal{R}+(1-\gamma)\,r\,\mathcal{R}\,\partial_r\,\mathcal{F}+2\,\mathcal{R}\,\mathcal{F}\big]}{\mathcal{R}^2\,\big[4\,r\,\mathcal{F}\,\partial_r\,\mathcal{R}+(1-2\,\gamma)\,r\,\mathcal{R}\,\partial_r\,\mathcal{F}+2\,\mathcal{R}\,\mathcal{F}\big]}.
\end{equation}
Eqs. \eqref{kh12_label} and \eqref{kh13_label} are complicated functions of $r$ and are difficult to investigate, so we reduce our calculations to a simpler case, where $ \gamma = 1 $. Therefore, Eqs. \eqref{kh12_label} and \eqref{kh13_label} are simplified as follows
\begin{equation}\label{kh14_label}
	\frac{L^2}{m_0^2}=\frac{\eta\,(m\,r-q_e^2)}{r^2-3\,m\,r+2\,q_e^2},
\end{equation}
\begin{equation}\label{kh15_label}
	\frac{E^2}{m_0^2}=\frac{\eta\,(r^2-2\,m\,r+q_e^2)^2}{r^2\,(r^2-3\,m\,r+2\,q_e^2)}.
\end{equation}
If the circular motion of neutral particles is timelike ($ \eta = 1 $), considering that the motion of particles outside the naked singularity is investigated and the positiveness of the left side of Eqs. \eqref{kh14_label} and \eqref{kh15_label}, we must have
\begin{equation}\label{kh16_label}
	\begin{aligned}
		&r>m+\sqrt{m^2-q_e^2},\quad\text{and}\quad
		r\ge\frac{q_e^2}{m},\quad\text{and}\\
		&r\le\frac{1}{2}\,\big(3\,m-\sqrt{9\,m^2-8\,q_e^2}\big),\quad\text{or}\quad r\ge\frac{1}{2}\,\big(3\,m+\sqrt{9\,m^2-8\,q_e^2}\big).
	\end{aligned} 
\end{equation}
In Fig. \ref{radi}, the graphs related to Eqs. \eqref{kh16_label} are depicted. According to this diagram, we obtain the minimum distance allowed for the circular motion of neutral particles.

\section{A Class of Charged Taub-NUT Metrics in the Presence of a Scalar Field}\label{CTNS}
The charged Taub-NUT metric can be regarded as a special case of the general Plebanski–Demianski solution presented in 1976 
\cite{plebanski1976rotating}. However, it was not until 2006 that this spacetime was explicitly formulated and systematically analyzed as an independent solution by Griffiths and Podolsk{\`y}
\cite{griffiths2006new}.
In this section, we are going to add a NUT parameter to metric \eqref{710c_label} and obtain the charged Taub-NUT metric in the presence of a scalar field. To do this, we will use the Ehlers transformations. To apply these transformations, we must first write the metric \eqref{710c_label} in the form of the LWP metric, i.e., Eq. \eqref{h1_label}. In this case, the relation between $ \rho $ and $ z $ with $ r $ and $ \theta $ is as follows
\begin{equation}\label{n1}
	\rho=\sqrt{r^2-2\,m\,r+q^2}\,\sin\theta,\qquad z=(r-m)\,\cos\theta,
\end{equation}
and the functions $ F $, $ \omega $, and $ e^{2\lambda} $ are characterized by the following equations
\begin{equation}\label{n2}
	\begin{aligned}
		&F=1-\frac{2\,m}{r}+\frac{q^2}{r^2},\\
		&\omega=0,\\
		&e^{2\lambda}=\Big(1+\frac{(m^2-q^2)\,\sin^2\theta}{r^2-2\,m\,r+q^2}\Big)^{\nu-1}.
	\end{aligned}
\end{equation}
The Ehlers transformations for the Ernst potentials $ \varepsilon $ and $ \Phi $ are as follows
\begin{equation}\label{eh1}
	\varepsilon^\prime=\frac{\varepsilon}{1+i\,c\,\varepsilon},\qquad
	\Phi^\prime=\frac{\Phi}{1+i\,c\,\varepsilon},
\end{equation}
where $ c $ is a constant and real parameter. Using Eqs. \eqref{h5_label}, \eqref{710c_label}, and \eqref{710cd_label}, we obtain the functions $ \varepsilon $, $ \Phi $, and $ \chi $ as follows
\begin{equation}\label{eh2}
	\begin{aligned}
		&\varepsilon=1-\frac{2\,m}{r},\\
		&\Phi=\frac{q}{r},\\
		&\chi=0.
	\end{aligned}
\end{equation}
Considering Eqs. \eqref{eh1} and \eqref{eh2}, we have:
\begin{equation}\label{eh3}
	\begin{aligned}
		&\varepsilon^\prime=\frac{r^2-2\,m\,r}{r^2+c^2\,(r-2\,m)^2}-i\,\frac{c\,(r-2\,m)^2}{r^2+c^2\,(r-2\,m)^2},\\
		&\Phi^\prime=\frac{q\,r}{r^2+c^2\,(r-2\,m)^2}-i\,\frac{c\,q\,(r-2\,m)}{r^2+c^2\,(r-2\,m)^2}.
	\end{aligned}
\end{equation}
According to Eqs. \eqref{h5_label} and \eqref{eh3}, we obtain the functions $ F^\prime $ and $ \chi^\prime $ as follows
\begin{equation}\label{eh4}
	\begin{aligned}
		&F^\prime=\text{Re}(\varepsilon^\prime)+\lvert\Phi^\prime\rvert^2=\frac{r^2-2\,m\,r+q^2}{r^2+c^2\,(r-2\,m)^2},\\
		&\chi^\prime=\text{Im}(\varepsilon^\prime)=-\frac{c\,(r-2\,m)^2}{r^2+c^2\,(r-2\,m)^2}.
	\end{aligned}
\end{equation}
Also, considering the function $ \Phi^\prime $ in Eq. \eqref{eh1}, the functions $ A_t^\prime $ and $ \tilde{A}_\phi^\prime $ are determined by the following equations
\begin{equation}\label{eh5}
	\begin{aligned}
		&A_t^\prime=\text{Re}(\Phi^\prime)=\frac{q\,r}{r^2+c^2\,(r-2\,m)^2},\\
		&\tilde{A}_\phi^\prime=\text{Im}(\Phi^\prime)=-\,\frac{c\,(r-2\,m)^2}{r^2+c^2\,(r-2\,m)^2}.
	\end{aligned}
\end{equation}
Now we want to obtain the function $ \omega^\prime $. To obtain it, we need to use Eq. \eqref{h6_label}. But as we mentioned before, the gradients of Eq. \eqref{h6_label} are written in the cylindrical coordinate system, and we use the following equation to write them in the spherical coordinate system in this case
\begin{equation}\label{eh6}
	\nabla\,h(r,\,\theta)=\frac{1}{\sqrt{(r-m)^2+(q^2-m^2)\,\cos^2\theta}}\;\big[\frac{\partial\,h(r,\,\theta)}{\partial\,r}\,\sqrt{r\,(r-2\,m)+q^2}\,\hat{r}+\frac{\partial\,h(r,\,\theta)}{\partial\,\theta}\,\hat{\theta}\,\big],
\end{equation}
So, using Eqs. \eqref{h6_label}, \eqref{n1}, \eqref{eh3}, \eqref{eh4}, and \eqref{eh6}, we obtain $ \omega^\prime $ as follows
\begin{equation}\label{eh7}
	\omega^\prime=4\,c\,m\,\cos\theta.
\end{equation}
It is recognized that relying solely on Ehlers coordinate transformations is insufficient for determining the vector potential. Since Ehlers transformations modify the orientation of the electromagnetic field, we will study a duality rotation applied to the Ernst potential in the following manner
\begin{equation}\label{eh8}
	\bar{\Phi}^\prime=\Phi^\prime\,e^{i\beta},
\end{equation}
In Eq. \eqref{eh8}, $ \beta $ is the rotation parameter. The components $ A_t^\prime $ and $ \tilde{A}_\phi^\prime $ in Eq. \eqref{eh5} are rotated as follows
\begin{equation}\label{eh9}
	\begin{pmatrix}
		\bar{A}_t^\prime\\
		\bar{\tilde{A}}_\phi^\prime\\
	\end{pmatrix}
	=
	\begin{pmatrix}
		\cos\beta&-\sin\beta\\
		\sin\beta&\cos\beta\\
	\end{pmatrix}
	\,
	\begin{pmatrix}
		A_t^\prime\\
		\tilde{A}_\phi^\prime\\
	\end{pmatrix}.
\end{equation}
Now, using Eqs. \eqref{eh5} and \eqref{eh9}, we have
\begin{equation}\label{eh10}
	\bar{A}_t^\prime=\frac{q\,\big[r\,\cos\beta+c\,(r-2\,m)\sin\beta\big]}{r^2+c^2\,(r-2\,m)^2},
\end{equation}
Also, using Eqs. \eqref{h7_label}, \eqref{eh1}, \eqref{eh4}, \eqref{eh5}, \eqref{eh6}, \eqref{eh7}, \eqref{eh9}, and \eqref{eh10}, we obtain $ \bar{A}_\phi^\prime $ as follows
\begin{equation}\label{eh11}
	\bar{A}_\phi^\prime=q\,\cos\theta\,(c\,\cos\beta-\sin\beta)-\omega^\prime\,\bar{A}_t^\prime.
\end{equation}
Ehlers transformations do not change the $ \lambda $ function like Harrison transformations ($ \lambda^\prime = \lambda $). So, by putting the functions $ \rho $, $ z $, $ e^{2\lambda} $, $ F^\prime $, and $ \omega^\prime $ from Eqs. \eqref{n1}, \eqref{n2}, \eqref{eh4}, and \eqref{eh7} into metric \eqref{h1_label}, we have
\begin{equation}\label{eh12}
	\begin{aligned}
		ds^{\prime^2}=&-\frac{r^2-2\,m\,r+q^2}{r^2+c^2\,(r-2\,m)^2}\,\big(dt-4\,c\,m\,\cos\theta\,d\phi\big)^2+\big[r^2+c^2\,(r-2\,m)^2\big]\,\Big(1+\frac{(m^2-q^2)\,\sin^\theta}{r^2-2\,m\,r+q^2}\Big)^\nu\\
		&\times\Big[\frac{dr^2}{r^2-2\,m\,r+q^2}+d\theta^2\Big]+\big[r^2+c^2\,(r-2\,m)^2\big]\,\sin^2\theta\,d\phi^2.
	\end{aligned}
\end{equation}
Now, using the following coordinate transformation
\begin{equation}\label{eh13}
	\begin{aligned}
		&\tilde{r}=\frac{1}{\sqrt{1+c^2}}\,\big[r\,(1+c^2)-2\,m\,c^2\big],\qquad\tilde{t}=\frac{t}{\sqrt{1+c^2}},\qquad\tilde{q}=q\,\sqrt{1+c^2},\\
		&m=-\frac{n}{2\,c}\,\sqrt{1+c^2},\qquad c=\frac{\tilde{m}-\sqrt{\tilde{m}^2+n^2}}{n},\qquad\cos\beta=\frac{1}{\sqrt{1+c^2}}.		
	\end{aligned}
\end{equation}
and removing the tilde from the parameters ($ \tilde{t} \to t $, $ \tilde{r} \to r $, $ \tilde{m} \to m $, and $ \tilde{q} \to q $), we write metric \eqref{eh12} as follows
\begin{equation}\label{eh14}
	ds^{\prime^2}=-\mathscr{F}\,[dt+\mathscr{W}\,d\phi]^2+\mathscr{K}^\nu\,\Big[\frac{dr^2}{\mathscr{F}}+(r^2+n^2)\,d\theta^2\Big]+(r^2+n^2)\,\sin^2\theta\,d\phi^2,
\end{equation}
where the functions $ \mathscr{F} $ and $ \mathscr{K} $ are as below
\begin{equation}\label{eh15}
	\begin{aligned}
		&\mathscr{F}=\frac{r^2-2\,m\,r+q^2-n^2}{r^2+n^2},\\
		&\mathscr{W}=2\,n\,\cos\theta,\\
		&\mathscr{K}=1+\frac{(m^2+n^2-q^2)\,\sin^2\theta}{r^2-2\,m\,r+q^2-n^2}.		
	\end{aligned}
\end{equation}
Considering Eqs. \eqref{eh7}, \eqref{eh10}, and \eqref{eh11}, and also using transformations \eqref{eh13}, we obtain the vector potential function $ A^\prime $ as follows
\begin{equation}\label{eh16}
	A^\prime=A_t^\prime\,dt+A_\phi^\prime\,d\phi=\frac{q\,r}{r^2+n^2}\,(dt+2\,n\,\cos\theta\,d\phi).
\end{equation}
Using $ \Box \varphi^\prime = 0 $ and metric \eqref{eh14}, we obtain the scalar field $ \varphi^\prime $ as follows
\begin{equation}\label{eh17}
	\varphi^\prime=\sqrt{-\frac{\nu}{2}}\,\ln\Big(\frac{r-m-\sqrt{m^2+n^2-q^2}}{r-m+\sqrt{m^2+n^2-q^2}}\Big).
\end{equation}
Metric \eqref{eh14} reduces to the charged Taub-NUT metric when $\nu = 0$
\cite{plebanski1976rotating, griffiths2006new}, 
to the Taub-NUT metric with a scalar field when $q = 0$
\cite{derekeh2024class}, 
and to the standard Taub-NUT metric when both $\nu = 0$ and $q = 0$
\cite{taub1951empty, newman1963empty, kramer1980exact}.
In the following, we will investigate the singularities of metrics \eqref{eh14}, and later, we will study some astrophysical aspects of these metrics.
\subsection{Singularities of a Class of Charged Taub-NUT Metrics in the Presence of a Scalar Field}
\begin{figure}[ht]
\centering
\includegraphics[width=0.6\linewidth]{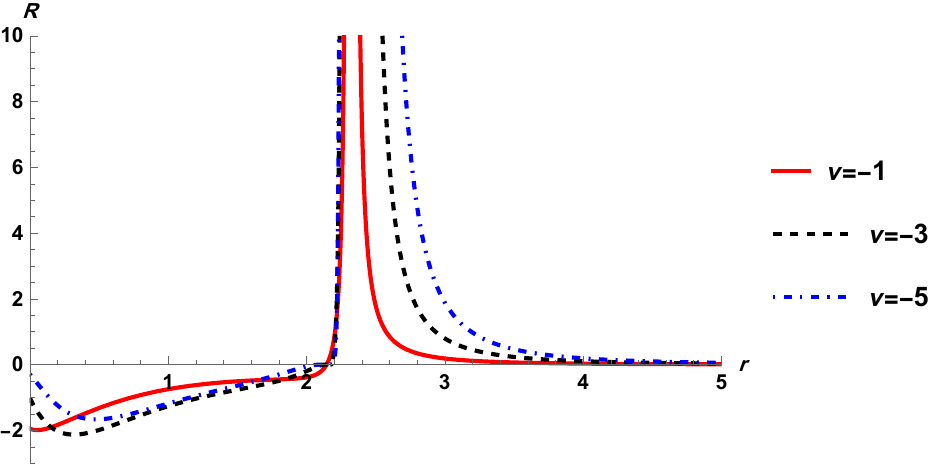}
\caption[]{\small The plot of the Ricci scalar in Eq.~\eqref{singctns} for the charged Taub-NUT metrics in the presence of a scalar field as a function of $r$ is shown for $m = 1$, $n = 1$, $q = 1/2$ and $\theta = \pi/6$. The diagrams show a singularity at point $m + \sqrt{m^2 + n^2 - q^2}$ and, as expected, the Ricci scalar approaches zero at large distances.}
\label{riccichTN}
\end{figure}

To investigate the singularities of the metrics \eqref{eh14}, we obtain the corresponding Ricci scalar as follows
\begin{equation}\label{singctns}
	R(r,\,\theta)=-\frac{2\,\nu\,(m^2-q^2+n^2)\,[r^2-2\,m\,r+q^2-n^2]^{\nu-1}}{(r^2+n^2)\,[r^2-2\,m\,r+m^2\,\sin^2\theta+(q^2-n^2)\,\cos^2\theta]^\nu}.
\end{equation}
According to Eq. \eqref{eh17}, $ \nu $ is always less than or equal to zero. Therefore, in Eq. \eqref{singctns} the metric singularity points are as follows
\begin{equation}\label{singctns2}
	r_\text{singularity}=m\pm\sqrt{m^2+n^2-q^2}.
\end{equation}
Therefore, metric \eqref{eh14} has a naked singularity at the two points specified in Eq. \eqref{singctns2}. In the following, we have depicted the Ricci scalar diagram as shown in Fig. \ref{riccichTN}.
\subsection{AXIS SINGULARITIES AT $\theta=0$ AND $\theta=\pi$}\label{axis-singularity-rewritten}

The general form of charged Taub-NUT metrics in the presence of a scalar field can be written as follows
\begin{equation}\label{rw1}
	ds^2 = -\mathscr{F}(r)\left(dt + 2n\cos\theta\,d\phi\right)^2 + \frac{\mathscr{K}(r,\theta)^\nu}{\mathscr{F}(r)}dr^2 + (r^2 + n^2)\left[\mathscr{K}(r,\theta)^\nu\,d\theta^2 + \sin^2\theta\,d\phi^2\right],
\end{equation}
where the functions $\mathscr{F}(r)$ and $\mathscr{K}(r,\theta)$ are defined by
\begin{equation}\label{rw2}
	\begin{aligned}
		&\mathscr{F}(r)=\frac{r^2-2\,m\,r+q^2-n^2}{r^2+n^2},\\
		&\mathscr{K}(r,\,\theta)=1+\frac{(m^2+n^2-q^2)\,\sin^2\theta}{r^2-2\,m\,r+q^2-n^2}.
	\end{aligned}
\end{equation}
When $\nu=0$, the scalar field vanishes and Eq. \eqref{rw1} reduces to the charged Taub-NUT metric. It's important to note that while the spacetime exhibits a singular behavior at $r = m \pm \sqrt{m^2 + n^2-q^2}$, it remains regular at $r = 0$. On the symmetry axis, however, i.e., at $\theta=0$ and $\theta=\pi$, the metric becomes singular due to the alteration of its signature, deviating from the usual $(-,+,+,+)$ structure.

To handle this axial singularity, consider the coordinate transformation $t = \bar{t} - 2n\phi$, which removes the singularity along $\theta=0$ and leads to the modified metric
\begin{equation}\label{rw3}
	\begin{aligned}
		ds^2 = -\mathscr{F}(r)\left(d\bar{t} + 2n(\cos\theta - 1)d\phi\right)^2 + \frac{\mathscr{K}(r,\theta)^\nu}{\mathscr{F}(r)}dr^2+ (r^2 + n^2)\left[\mathscr{K}(r,\theta)^\nu\,d\theta^2 + \sin^2\theta\,d\phi^2\right].
	\end{aligned}
\end{equation}
This adjustment makes the axis at $\theta=0$ regular, allowing us to interpret $\phi$ as a periodic coordinate ranging from $0$ to $2\pi$. However, the same singular behavior remains at $\theta=\pi$, manifesting as a conical defect.

An alternative transformation, $t = \bar{t} + 2n\phi$, regularizes the axis at $\theta=\pi$ instead
\begin{equation}\label{rw3b}
	\begin{aligned}
		ds^2 = -\mathscr{F}(r)\left(d\bar{t} + 2n(\cos\theta + 1)d\phi\right)^2 + \frac{\mathscr{K}(r,\theta)^\nu}{\mathscr{F}(r)}dr^2+ (r^2 + n^2)\left[\mathscr{K}(r,\theta)^\nu\,d\theta^2 + \sin^2\theta\,d\phi^2\right].
	\end{aligned}
\end{equation}
Despite this, a singularity still persists at $\theta=0$ in this new frame.

There exist differing viewpoints on the nature of this axial singularity. While Bonnor \cite{bonnor1969new} interprets it as physically meaningful and seeks its physical origins, Misner proposes eliminating the singularity by introducing closed timelike curves (CTCs) \cite{misner1963flatter, misner1967contribution}, a concept further discussed in the following section.

It is evident from both Eq. \eqref{rw3} and Eq. \eqref{rw3b} that the addition of the scalar field does not influence the singularity structure at $\theta=0$ and $\theta=\pi$, implying that the physical interpretations valid for the charged Taub-NUT spacetime remain applicable to the charged Taub-NUT metrics in the presence of a scalar field.

\subsection{CLOSED TIMELIKE CURVES IN THE CHARGED TAUB-NUT METRICS IN THE PRESENCE OF A SCALAR FIELD}\label{CTC-rewritten}

Solutions containing closed timelike curves (CTCs) are well-known within general relativity \cite{tipler1976causality, visser1995lorentzian, lobo2010closed}. Such curves allow an observer to revisit past events, challenging the causality principle. A key consequence of their presence is the tilting of the light cones.

The charged Taub-NUT metric is among the spacetimes permitting CTCs. We now analyze whether the scalar field introduced in charged Taub-NUT metrics affects their presence.

Using the metrics in Eqs. \eqref{rw3} and \eqref{rw3b}, consider a circular curve given by:
\[
\tau = (t = \text{const},\, r = \text{const},\, \theta = \text{const}).
\]
The proper length of this curve is obtained as:
\begin{equation}\label{rw5}
	s_\tau^2 = (2\pi)^2\left[-\frac{4n^2(r^2 - 2mr +q^2 - n^2)\cos^2\theta}{r^2 + n^2} + (r^2 + n^2)\sin^2\theta\right] = (2\pi)^2\,Z(r,\theta).
\end{equation}

This curve becomes a closed timelike curve if $Z(r,\theta)<0$, and it corresponds to a closed null curve if $Z(r,\theta)=0$.

As seen in this derivation, the scalar field introduced in a class of charged Taub-NUT metrics does not influence the existence or structure of the CTCs. Hence, the results and interpretations derived for CTCs in the Taub-NUT geometry are also valid in the charged Taub-NUT metrics in the presence of a scalar field.
\section{Gravitational lensing}\label{5}
Since gravity curves spacetime, light, like ordinary matter, is affected by it and deflected in a gravitational field. When photons approach a massive object, their path is bent. In this study, we aim to calculate the deflection angle of light in the gravitational fields of the metrics \eqref{1_label} and \eqref{eh14}, and explore how scalar and electric fields influence the deflection angle of light.
\subsection{Gravitational Lensing for a Class of Charged Metrics in the Presence of a Scalar Field}\label{5a1}
In this subsection, we aim to calculate the deflection angle of light in gravitational lensing for the metrics \eqref{1_label}. First, we consider the optical metric in the equatorial plane ($ \theta = \pi / 2 $) of spacetime \eqref{1_label} 
\begin{equation}\label{49_label}
	d t^2 = \frac{g_{r r}}{g_{tt}} \, d r^2 + \frac{g_{\phi \phi}}{g_{tt}} \, d\phi^2,
\end{equation}
where,
\begin{equation}\label{50_label}
	\begin{aligned}
		& \frac{g_{r r}}{g_{tt}} = \mathcal{R}^4 \, \mathcal{F}^{\mu - 1 - \gamma} \; \mathcal{K}^{\nu},\\
		& \frac{g_{\phi \phi}}{g_{tt}} = r^2 \, \mathcal{R}^4 \, \mathcal{F}^{1 - 2 \, \gamma}.
	\end{aligned}
\end{equation}
The equation for the Gaussian curvature is as follows
\begin{equation}\label{51_label}
	\mathfrak{K}= - \, \frac{1}{\sqrt{g}} \, \Big[ \partial_r \, \Big( \sqrt{\frac{g_{tt}}{g_{rr}}} \, \partial_r \, \sqrt{\frac{g_{\phi\phi}}{g_{tt}}}\, \Big) + \partial_\phi \, \Big( \sqrt{\frac{g_{tt}}{g_{\phi\phi}}} \, \partial_\phi \, \sqrt{\frac{g_{rr}}{g_{tt}}}\, \Big) \Big],
\end{equation}
where $ \sqrt{g} $ is equal to
\begin{equation}\label{52_label}
	\sqrt{g} = r \, \mathcal{R}^4 \, \mathcal{F}^{\frac{1}{2} \, (\mu - 3\, \gamma)} \; \mathcal{K}^\frac{\nu}{2}.
\end{equation}
The total deflection angle in gravitational lensing is calculated as follows
\begin{equation}\label{53_label}
	\delta = - \, \int_{0}^{\pi} \, \int_{r_0}^{\infty} \, \mathfrak{K} \, dS,
\end{equation}
where $ r_0 $ is the minimum distance from the source, and $ dS $ is defined as follows
\begin{equation}\label{54_label}
	d S = \sqrt{g} \, d r \, d \phi.
\end{equation}
By substituting Eqs. (\ref{50_label}–\ref{52_label}) and Eq. \eqref{54_label} into Eq. \eqref{53_label}, we have
\begin{equation}\label{55_label}
	\delta = - \, \int_{0}^{\pi} \, \int_{r_0}^{\infty} \, \partial_r \, \big[ \mathcal{F}^{1 - \frac{1}{2} \, (\gamma + \mu)} \; \mathcal{K}^{- \, \frac{\nu}{2}} \, \big( 1 + 2 \, r \, \frac{\mathcal{R}^\prime}{\mathcal{R}} + r \, \big( \frac{1}{2} - \gamma \big) \, \frac{\mathcal{F}^\prime}{\mathcal{F}} \big) \big] \, d r \, d \phi.
\end{equation}
By using Eqs. \eqref{3_label} in Eq. \eqref{55_label} and expanding the terms up to order $ r^{-2} $, we have
\begin{equation}\label{57_label}
	\begin{aligned}
		\delta=&-\int_{0}^{\pi} \, \int_{r_0}^{\infty} \,\partial_r\,\big\{1-\frac{2\,m\,\gamma}{r}+\frac{1}{2\,r^2}\,\big[(q_e^2+q_m^2)\big(4\,\mu^2+8\,\mu\,(\nu-1)+\nu\,(4\,\nu-7)+3\big)\\
		&+m^2\,(4\,\mu+3\,\nu-3)\big]\big\}\, d r \, d \phi.
	\end{aligned}
\end{equation}
Now, we need to determine the range of $ r_0 $. For this purpose, we first consider the metric \eqref{1_label} for $ \theta = \pi / 2 $ as follows
\begin{equation}\label{58_label}
	ds^2 = - \frac{\mathcal{F}^\gamma}{\mathcal{R}^2} \, dt^2 + \frac{\mathcal{R}^2 \, \mathcal{F}^\mu \, \mathcal{K}^\nu}{\mathcal{F}} \, dr^2 + r^2 \, \mathcal{R}^2 \, \mathcal{F}^{1 - \gamma} \, d \phi^2.
\end{equation}
We consider the geodesic Lagrangian as follows
\begin{equation}\label{60_label}
	\mathcal{L} = - \frac{1}{2} \, \frac{\mathcal{F}^\gamma}{\mathcal{R}^2} \, \dot{t}^2 + \frac{1}{2} \, \frac{\mathcal{R}^2 \, \mathcal{F}^\mu \, \mathcal{K}^\nu}{\mathcal{F}} \, \dot{r}^2 + \frac{1}{2} \, r^2 \, \mathcal{R}^2 \, \mathcal{F}^{1 - \gamma} \, \dot{\phi}^2,
\end{equation}
The dot represents $ \frac{d}{d\tau} $, where $ \tau $ is the affine parameter. The Lagrangian equations of motion are as follows
\begin{equation}\label{61_label}
	E =\frac{\mathcal{F}^\gamma}{\mathcal{R}^2} \, \frac{d t}{d \tau} =  const,
\end{equation}
\begin{equation}\label{62_label}
	L = \mathcal{R}^2 \, \mathcal{F}^{1 - \gamma} \, r^2 \, \frac{d \phi}{d \tau} = const.
\end{equation}
\begin{equation}\label{63_label}
	\frac{E}{L} =\frac{\mathcal{F}^\gamma}{\mathcal{R}^4 \, \mathcal{F}^{1 - \gamma} \, r^2} \, \frac{d t}{d \phi} = \frac{1}{b},
\end{equation}
where $ b $ is a constant parameter. Also, according to Eq. \eqref{58_label}, $ \left( \frac{dr}{d\tau} \right)^2 $ is equal to
\begin{equation}\label{64_label}
	\frac{\mathcal{R}^2 \, \mathcal{F}^\mu \, \mathcal{K}^\nu}{\mathcal{F}} \, \big( \frac{d r}{d \tau} \big)^2 = \frac{\mathcal{F}^\gamma}{\mathcal{R}^2} \, \big( \frac{d t}{d \tau} \big)^2 - r^2 \, \mathcal{R}^2 \, \mathcal{F}^{1 - \gamma} \, \big( \frac{d \phi}{d \tau} \big)^2.
\end{equation}
Using Eq. \eqref{63_label} and $ d\phi $ instead of $ d\tau $ in Eq. \eqref{64_label}, we have
\begin{equation}\label{65_label}
	\frac{\mathcal{R}^2 \, \mathcal{F}^\mu \, \mathcal{K}^\nu}{\mathcal{F}} \, \big( \frac{d r}{d \phi} \big)^2 = \frac{r^4 \, \mathcal{R}^6 \, \mathcal{F}^{2\,(1 - \gamma)}}{b^2 \, \mathcal{F}^\gamma} - r^2 \, \mathcal{R}^2 \, \mathcal{F}^{1 - \gamma}.
\end{equation}
By defining $ u = 1/r $, we can rewrite Eq. \eqref{65_label} as follows
\begin{equation}\label{67_label}
	\big( \frac{d u}{d \phi} \big)^2 = \frac{\mathcal{F}^{2 - \gamma}}{\mathcal{F}^\mu \, \mathcal{K}^\nu} \, \big( \frac{\mathcal{R}^4 \,\mathcal{F}^{1 -2\, \gamma}}{b^2} - u^2\big).
\end{equation}
By differentiating Eq. \eqref{67_label} with respect to $ u $ and expanding the resulting expression up to the order of $ u^2 $, we have
\begin{equation}\label{68_label}
	\begin{aligned}
		\frac{d^2 u}{d \phi^2} + u =& 3 \, m \, u^2-\frac{2\,m\,(\mu+\nu)}{b^2}+\frac{m^2\,u}{b^2}\,\Big\{p^2\,\big[4\,\mu\,(\nu-1)+\nu\,(2\,\nu-5)+2\,\mu^2\big]+6\,(\mu+\nu)^2\Big\}\\
		&-\frac{m^3\,u^2}{b^2}\,\Big\{p^2\,\big[\mu\,\big(6\,\nu\,(5\,\nu-9)+8\big)+\nu\,\big(10\,\nu\,(\nu-3)+11\big)+6\,\mu^2\,(5\,\nu-4)\\
		&+10\,\mu^3\big]+6\,(\mu+\nu)^3\Big\}.
	\end{aligned}
\end{equation}
First, we solve the following homogeneous equation
\begin{equation}\label{69_label}
	\frac{d^2 u}{d \phi^2} + u = 0.
\end{equation}
The general solution to Eq. \eqref{69_label} can be written as follows
\begin{equation}\label{70_label}
	u = \frac{\sin \phi}{b}.
\end{equation}
In Eq. \eqref{70_label}, $ b $ is the impact parameter. Now, by using Eq. \eqref{70_label} in Eq. \eqref{68_label} and keeping terms up to order $ b^{-2} $, we have
\begin{equation}\label{71_label}
	\frac{d^2 u}{d \phi^2} + u - 3 \, m \, u^2 + \frac{2\,m\,(\mu+\nu)}{b^2} = 0.
\end{equation}
The general solution to Eq. \eqref{71_label} is as follows
\begin{equation}\label{72_label}
	u = \frac{\sin \phi}{b} + \frac{m}{b^2} \, \big[1 + \cos^2 \phi-2\,(\mu+\nu)\big].
\end{equation}
Eqs. \eqref{67_label} and \eqref{72_label} are consistent up to order $ b^{-2} $. We can substitute the solution from Eq. \eqref{72_label} into Eq. \eqref{57_label} and obtain the following equation
\begin{equation}\label{73_label}
	\begin{aligned}
		\delta=&\frac{4\,\gamma\,m}{b}+\frac{\pi\,\gamma\,m^2\,(4\,\gamma-1)}{b^2}-\frac{\pi}{4\,b^2}\,\Big\{q_e^2\,\big[4\,\big(2\,\mu\,(\nu-1)+\mu^2+\nu^2\big)-7\,\nu+3\big]\\
		&+m^2\,(\mu-3\,\gamma)\Big\}.
	\end{aligned}
\end{equation}
If we choose $ \mu = \gamma^2 - \gamma $ and $ \nu = 1 - \gamma^2 $, the $ \delta $ for the ZV metric is obtained as follows
\begin{equation}\label{74a_label}
	\delta_\gamma=\frac{4\,\gamma\,m}{b}+\frac{15\,\pi\,\gamma^2\,m^2}{4\,b^2}-\frac{3\,\pi\,\gamma^2\,q_e^2}{4\,b^2}.
\end{equation}
The deflection angle of light for the FJNW metric ($ \mu = 1 - \gamma, , \nu = 0 $) is equal to
\begin{equation}\label{74_label}
	\delta_{FJNW}=\frac{4\,\gamma\,m}{b}+\frac{\pi\,m^2\,(16\,\gamma^2-1)}{4\,b^2}+\frac{\pi\,(1-4\,\gamma^2)\,q_e^2}{4\,b^2}.
\end{equation}
By investigating the geodesics related to the metrics \eqref{1_label}, we found that $ \gamma $ is always greater than or equal to $ \frac{1}{2} $. Considering this and according to Eqs. \eqref{74a_label} and \eqref{74_label}, it is clear that the presence of electric charge in the metrics causes the deflection angle of light to decrease.

The difference in the deflection angle of light in the background of the charged FJNW and charged ZV metrics is as follows
\begin{equation}\label{74b_label}
	\delta_\gamma-\delta_{FJNW} =\frac{\pi\,(m^2-q_e^2)\,(1-\gamma^2)}{4\,b^2}.
\end{equation}
Considering $ \lvert q_e \rvert \le m $ and $ \gamma > \frac{1}{2} $, the different cases of Eq. \eqref{74b_label} are as follows
\begin{equation}\label{74abc_label}
	\delta_\gamma-\delta_{FJNW}=\begin{cases}
		>0,\qquad\,  \text{for}\; \frac{1}{2}<\gamma<1,\\0,\quad\qquad\,  \text{for}\; \gamma=1,\\<0,\qquad\,  \text{for}\; \gamma>1.
	\end{cases}
\end{equation}
Therefore, according to Eq. \eqref{74abc_label}, the presence of a scalar field can either increase or decrease the deflection angle of light, depending on the value of $ \gamma $.

By choosing $ \gamma = 1 $ and $ \mu = -\nu $ in Eq. \eqref{73_label}, we obtain the deflection angle related to the RN metric as follows
\begin{equation}\label{75_label}
	\delta_{RN} = \frac{4 \, m}{b}+\frac{15\,\pi\,m^2}{4\,b^2}-\frac{3\,\pi\,q_e^2}{4\,b^2},\qquad\delta_S = \frac{4 \, m}{b} + \frac{15 \, \pi\,m^2}{4 \, b^2}.
\end{equation}
where $ \delta_S $ represents the deflection angle of light in the Schwarzschild metric. It is also evident from Eqs. \eqref{75_label} that the electric charge causes the deflection angle to decrease.
\subsection{Gravitational Lensing in Charged Taub-NUT Metrics with a Scalar Field}
In this subsection, we aim to derive the light deflection angle for gravitational lensing in the metrics given by \eqref{eh14}. To achieve this, we follow the same approach as in subsection \ref{5a1} and calculate the deflection angle for a class of charged Taub-NUT metrics in the presence of a scalar field, as outlined below
\begin{equation}\label{dactns_label}
	\delta_{CTNS}=\frac{4 \, m}{b}+\frac{\pi\,m^2\,(\nu+15)}{4\,b^2}-\frac{\pi\,q^2\,(\nu+3)}{4\,b^2}+\frac{\pi\,n^2\,(\nu+7)}{4\,b^2}.
\end{equation}
Now, we examine the impact of the NUT parameter on the light deflection angle in the metrics given by \eqref{eh14}. To do so, we consider three distinct cases:
\begin{itemize}
	\item 
	1) If $ \nu > -7 $, the presence of the NUT parameter results in an increase in the light deflection angle.
	\item 
	2) If $ \nu = -7 $, the presence of the NUT parameter has no effect on the deflection angle.
	\item 
	3) If $ \nu < -7 $, the NUT parameter decreases the light deflection angle.
\end{itemize}
The presence of the charge $ q $, depending on the value of $ \nu $, can either increase (for $ \nu < -3 $) or decrease (for $ \nu > -3 $) the light deflection angle, or have no effect on it (for $ \nu = -3 $).
\section{Quasi normal modes}\label{6a}
In this section, we aim to obtain the QNMs associated with the metrics given by \eqref{1_label} and \eqref{eh14} using the light ring method. For more details on the light ring method, refer to Refs. \cite{ferrari1984oscillations, ferrari1984new, mashhoon1985stability}.
\subsection{QNMs of the Reissner-Nordstr\"{o}m (RN) metric in the presence of a scalar field}\label{6b}
We first set $ \theta = \pi/2 $. For simplicity, we assume that the scalar field and electric charge are small. Thus, we expand the metric components up to first order in $ \nu $ and second order in $ q $, and omit terms proportional to $ \nu q $. The metric can then be written as follows
\begin{equation}\label{q1_label}
	\begin{aligned}
		ds^2&=-\big(1-\frac{2\,m}{r}+\frac{q^2}{r^2}\big)\,dt^2+\frac{r}{r-2\,m}\,\big[1+\nu\,\ln(1+\frac{m^2}{r^2-2\,m\,r})-\frac{q^2}{r\,(r-2\,m)}\big]\,dr^2\\
		&+r^2\,\big[1+\nu\,\ln(1+\frac{m^2}{r^2-2\,m\,r})\big]\,d\theta^2+r^2\,d\phi^2.
	\end{aligned}
\end{equation}
We apply the same approximations in the following calculation of the QNMs. Defining the QNMs as $ Q = \Omega + i , \tilde{\Gamma} $, we first focus on obtaining the real part of the QNMs. In the eikonal limit, using the light ring method, $ \Omega $ can be written as follows
\begin{equation}\label{q2_label}
	\Omega=\pm\,j\,\Omega_{\pm},	
\end{equation}
where $ \Omega_{\pm} $ is the frequency corresponding to the null rays of periodic orbits, and $ j $ is a real number. Perturbations of massless waves in an axisymmetric spacetime can be expressed as the following eigenvalue superpositions
\begin{equation}\label{qnms1_label}
	e^{i \, (\Omega \, t - \iota \, \phi)} \, S_{\Omega \, j \, \iota \, s} (r, \theta).
\end{equation}
where $ S $ and $ \Omega $ are the spin and frequency of the waves, respectively, and $ \iota $ and $ j $ are the angular momentum parameters that satisfy
\begin{equation}\label{qnms2_label}
	\lvert \iota \rvert \le j.
\end{equation}
Additionally, the following conditions hold in the eikonal limit
\begin{equation}\label{qnms3_label}
	\Omega \gg 1/M,\qquad\lvert \iota \rvert = j \gg 1.
\end{equation}
From the calculations of QNMs associated with different black holes, it is determined that the frequency of perturbations ($ \Omega $) is given by
\begin{equation}\label{qnms4_label}
	\Omega = \iota \, \frac{d \phi}{d t} = \pm \, j \, \Omega_{\pm}.
\end{equation} 
To calculate $ \Omega_{\pm} $, we first need to determine the radius of the null geodesic circles. For this, we use the following relations
\begin{equation}\label{q3_label}
	g_{t t} + g_{\phi \phi} \, \big( \frac{d \phi}{d t} \big)^2 = 0,
\end{equation}
\begin{equation}\label{q4_label}
	\Gamma_{t t}^r + \Gamma_{\phi \phi}^r  \, \big( \frac{d \phi}{d t} \big)^2 = 0.
\end{equation}
Equation \eqref{q3_label} holds for any null path, and equation \eqref{q4_label} represents the geodesic equation for the radial component. Using the metric in \eqref{q1_label}, equations \eqref{q3_label} and \eqref{q4_label} can be expressed in the following forms, respectively
\begin{equation}\label{q5_label}
	\frac{d\phi}{dt}=\sqrt{\frac{r-2\,m}{r^3}}\,\big( \frac{q^2}{2\,r^2\,(r-2\,m)}+1\big),
\end{equation}
\begin{equation}\label{q6_label}
	\frac{d\phi}{dt}=\sqrt{\frac{m}{r^3}}\,\big( 1-\frac{q^2}{2\,m\,r^2}\big).
\end{equation}
We now apply the following perturbation 
\begin{equation}\label{q7_label}
	r = 3 \, m + \bar{\epsilon}.
\end{equation}
In equation \eqref{q7_label}, $ 3,m $ represents the radius of the light ring of the Schwarzschild black hole. First, we substitute the value of $ r $ from equation \eqref{q7_label} into equations \eqref{q5_label} and \eqref{q6_label}. Then, we set the right-hand side of equation \eqref{q5_label} equal to the right-hand side of equation \eqref{q6_label}, which leads to the following equation
\begin{equation}\label{q7a_label}
	\bar{\epsilon}=- \frac{12\,m\,q^2}{18\,m^2-13\,q^2}.
\end{equation}
Now, using Eqs. \eqref{q7_label} and \eqref{q7a_label}, we can obtain the radius of the black hole's light ring as follows
\begin{equation}\label{q8_label}
	r_0 = 3\,m - \frac{12\,m\,q^2}{18\,m^2-13\,q^2}.
\end{equation}
By substituting $ r_0 $ into either equation \eqref{q5_label} or \eqref{q6_label}, we obtain
\begin{equation}\label{q9_label}
	\Omega_\pm=\frac{d\phi}{dt}=\frac{1}{3\,\sqrt{3}\,m}\,\big(1\pm\frac{q^2}{6\,m^2}\big).
\end{equation}
Now, we will calculate $ \tilde{\Gamma} $, which represents the divergence of null rays or the decay rate of the amplitude. To do this, we introduce a small perturbation to the coordinates $ x^\mu = (t, r, \theta, \phi) = (t, r_0, \frac{\pi}{2}, \Omega_\pm t) $, thereby perturbing the null circular orbit in the equatorial plane
\begin{equation}\label{q10_label}
	r = r_0 \, [1 + \epsilon_0 \, f_r(t)], \quad \phi = \Omega_\pm \, [t + \epsilon_0 \, f_\phi(t)], \quad \ell = t + \epsilon_0 \, f_\ell(t).
\end{equation}
In Eq. \eqref{q10_label}, $ \epsilon_0 $ is a small perturbation parameter, and $ f_r(t) $ and $ f_\ell(t) $ are functions that vanish at $ t = 0 $. From Eqs. \eqref{qnms4_label} and \eqref{q10_label}, it is clear that $ f_\phi(t) = 0 $. Additionally, we can express the perturbed propagation vector as follows
\begin{equation}\label{q11_label}
	K^\mu = \frac{d x^\mu}{d \ell} = \big(1 - \epsilon_0 \, \frac{\partial\,f_\ell}{\partial\,t}, \, \epsilon_0 \, r_0 \, \frac{\partial\,f_r}{\partial\,t}, \, 0, \, \Omega_\pm \, (1 - \epsilon_0 \, \frac{\partial\,f_\ell}{\partial\,t})\big).
\end{equation}
On the other hand, the conservation law for the convergence of null rays can be written as follows
\begin{equation}\label{q12_label}
	\nabla_\mu \, (\rho_n \, K^\mu) = 0,
\end{equation}
where $ \rho_n $ represents the density of null rays. Additionally, from Eq. \eqref{q12_label}, we have
\begin{equation}\label{q13_label}
	\frac{1}{\rho_n} \, \frac{d \rho_n}{d \ell} = - \nabla_\mu K^\mu = - \frac{1}{\sqrt{- g}} \, \frac{\partial}{\partial x^\alpha} \, (\sqrt{- g} \, K^\alpha),
\end{equation}
where $ g $ is the determinant of the metric in Eq. \eqref{q1_label}. Using Eqs. \eqref{q11_label} and \eqref{q13_label}, we have
\begin{equation}\label{65a-label}
	\begin{aligned}
		\frac{1}{\rho_n}\,\frac{d\rho_n}{d\ell}=&-\frac{1}{\sqrt{-g}}\,\frac{\partial}{\partial t}\,\big[\sqrt{-g}\,(1-\epsilon_0\,f_\ell^\prime)\big]-\frac{1}{\sqrt{-g}}\,\frac{\partial}{\partial r}\,\big[\sqrt{-g}\,\epsilon_0\,r_0\,f_r^\prime\big]\\
		&-\frac{\Omega_\pm}{\sqrt{-g}}\,\frac{\partial}{\partial\phi}\,\big[\sqrt{-g}\,(1-\epsilon_0\,f_\ell^\prime)\big].
	\end{aligned}
\end{equation}
According to the metric in Eq. \eqref{q1_label}, we obtain $ \sqrt{-g} $ as follows 
\begin{equation}\label{65ba-label}
	\sqrt{-g}=r^2+\nu\,r^2\,\ln\big(1+\frac{m^2}{r^2-2\,m\,r}\big).
\end{equation}
Considering Eqs. \eqref{q8_label} and \eqref{q10_label}, $ \sqrt{-g} $ can be expressed as follows
\begin{equation}\label{65b-label}
	\sqrt{-g}=9\,m^2\,(1+2\,\epsilon_0\,f_r+\nu\,\ln\frac{4}{3})-4\,q^2.
\end{equation}
Using Eq. \eqref{65b-label} in Eq. \eqref{65a-label}, we obtain
\begin{equation}\label{65c-label}
	\frac{1}{\rho_n} \, \frac{d \rho_n}{d \ell} = - \epsilon_0\,\big[\frac{\partial(2\,f_r-f_\ell^\prime)}{\partial t}+\frac{\partial(r_0\,f_r^\prime)}{\partial r}+\Omega_\pm\,\frac{\partial(2\,f_r-f_\ell^\prime)}{\partial\phi}\big].
\end{equation}
By using Eq. \eqref{q10_label} ($ dr/dt = \epsilon_0,r_0,f_r^\prime $) in Eq. \eqref{65c-label}, we obtain
\begin{equation}\label{65d-label}
	\frac{1}{\rho_n} \, \frac{d \rho_n}{d t} =\frac{d\,(\epsilon_0\,r_0\,f_r^\prime)}{dt}\,\frac{dt}{dr}+\mathcal{O}(\epsilon_0)= - \frac{f_r^{\prime \prime}(t)}{f_r^\prime(t)}+\mathcal{O}(\epsilon_0).
\end{equation}
To calculate $ \rho_n $, we first need to determine the function $ f_r(t) $. For this, we use the radial component of the geodesic equation
\begin{equation}\label{q15_label}
	\frac{d^2 r}{d \ell^2} + \Gamma_{t t}^r  \,\big( \frac{d t}{d \ell} \big)^2 + \Gamma_{\phi \phi}^r \, \big( \frac{d \phi}{d \ell} \big)^2 + \Gamma_{\theta \theta}^r \, \big( \frac{d \theta}{d \ell} \big)^2 = 0.
\end{equation}
By using Eqs. \eqref{q8_label} and \eqref{q10_label} in Eq. \eqref{q15_label}, expanding the result to the order of $ \epsilon_0 $, and setting $ \theta = \pi/2 $, we obtain
\begin{equation}\label{q16_label}
	\frac{\epsilon_0}{3\,m}\,(9\,m^2-2\,q^2)\,f_r^{\prime \prime}(t)+\frac{\epsilon_0}{81\,m^3}\,\big[q^2+9\,m^2\,(\nu\,\ln\frac{4}{3}-1)\big]\,f_r(t)=0.
\end{equation}
Therefore, we can express $ f_r(t) $ as
\begin{equation}\label{q17_label}
	f_r(t) = \sinh (\zeta\, t),
\end{equation}
where $ \zeta $ is
\begin{equation}\label{q18_label}
	\zeta = \frac{1}{54 \, \sqrt{3} \, m^3} \, \big[ q^2-9\,m^2\,(\nu\,\ln\frac{4}{3}-2) \big].
\end{equation}
Also, using Eqs. \eqref{65d-label} and \eqref{q17_label}, we can express $ \rho_n(t) $ as follows
\begin{equation}\label{q19_label}
	\rho_n(t) = \rho_n(0) \, \frac{1}{\cosh(\zeta)} \simeq 2 \, \rho_n(0) \, (e^{- \zeta \, t} - e^{- 3 \, \zeta \, t} + e^{- 5 \, \zeta \, t} - \dots).
\end{equation}
Thus, the imaginary part of the QNMs is given by
\begin{equation}\label{q20_label}
	\tilde{\Gamma} = \big( n + \frac{1}{2} \big) \, \zeta.
\end{equation}
Finally, the QNMs for the metric in Eq. \eqref{q1_label} are given by
\begin{equation}\label{q21_label}
	\begin{aligned}
		Q & = \Omega + i \, \tilde{\Gamma}\\
		& = j \, \big[ \frac{1}{3\,\sqrt{3}\,m}\,\big(1\pm\frac{q^2}{6\,m^2}\big) \big] + i \, \big( n + \frac{1}{2} \big) \, \Big\{ \frac{1}{54 \, \sqrt{3} \, m^3} \, \big[ q^2-9\,m^2\,(\nu\,\ln\frac{4}{3}-2) \big]\Big\}.
	\end{aligned}
\end{equation}
Eq. \eqref{q21_label}, in the limit $ q = \nu = 0 $, gives the QNMs of the Schwarzschild metric.

According to Eq. \eqref{q21_label}, when $ \gamma = 1 $, the presence of the scalar field does not affect the real part, i.e., the frequency of the QNMs. However, due to the negativity of the $ \nu $ parameter, it increases the imaginary part of the QNMs. Depending on the sign of the electric charge, it can either increase or decrease the frequency of the QNMs. The effect of the electric charge on the imaginary part of the QNMs is similar to that of the scalar field.
\subsection{QNMs of the charged FJNW metric ($ \nu = 0 $)}\label{6c}
Now, we derive the QNMs of the metric in Eq. \eqref{1_label} for $ \theta = \pi/2 $, $ \nu = 0 $, $ \mu = 1 - \gamma $, and $ \gamma = 1 + \xi $, up to first order in $ \xi $ and second order in $ q_e $ and $ q_m $. With these conditions, the metric in Eq. \eqref{1_label} can be written as follows
\begin{equation}\label{jnw1_label}
	\begin{aligned}
		ds^2&=-\big[1-\frac{2\,m}{r}+\frac{q_e^2+q_m^2}{r^2}+\xi\,(1-\frac{2\,m}{r})\,\ln\big(1-\frac{2\,m}{r}\big)\big]\,dt^2+\frac{1}{1-2\,m/r}\\
		&\times\big[1-\frac{q_e^2+q_m^2}{r^2\,(1-2\,m/r)}-\xi\,\ln\big(1-\frac{2\,m}{r}\big)\big]\,dr^2+r^2\,\big[1-\xi\,\ln\big(1-\frac{2\,m}{r}\big)\big]\,\big(d\theta^2+d\phi^2\big).
	\end{aligned}
\end{equation}
We can perform the same calculation as in subsection \ref{6b} to obtain the QNMs of the metric in Eq. \eqref{jnw1_label} as follows
\begin{equation}\label{jnw2_label}
	\begin{aligned}
		Q & = \Omega + i \, \tilde{\Gamma}\\
		&=j \, \big[ \frac{1}{3\,\sqrt{3}\,m}\,\big(1\mp\xi\,\ln3\pm\frac{q_e^2+q_m^2}{6\,m^2}\big) \big] + i \, \big( n + \frac{1}{2} \big) \, \Big\{ \frac{1}{3 \, \sqrt{3} \, m} \, \big[ 1-\xi\,\ln3+\frac{q_e^2+q_m^2}{18\,m^2} \big]\Big\}.
	\end{aligned}
\end{equation}
According to Eq. \eqref{jnw2_label}, the presence of the $ \xi $ parameter causes a shift in both the frequency (the real part of Eq. \eqref{jnw2_label}) and the imaginary part of the QNMs. The direction of this shift (increase or decrease) depends on the sign of $ \xi $ and the sign in Eq. \eqref{jnw2_label}. The presence of electric and magnetic charges increases the imaginary part of the QNMs, and depending on the sign in Eq. \eqref{jnw2_label}, it can either increase or decrease the frequency of the QNMs.
\subsection{QNMs of a class of charged Taub-NUT metrics in the presence of a scalar field}
Here, we aim to obtain the QNMs of the metrics in Eq. \eqref{eh14} using the method introduced in subsection \ref{6b}. To compute the QNMs, we first expand metric \eqref{eh14} to first order in $ \nu $ and second order in $ q $ and $ n $, while ignoring terms of order $ \nu,q $, $ \nu,n $, $ q^2,n $, and $ q,n^2 $
\begin{equation}\label{qnmsctns}
	\begin{aligned}
		ds^2=&-\big(1-\frac{2\,m}{r}+\frac{q^2}{r^2}+\frac{2\,n^2}{r^2}\,[\frac{m}{r}-1]\big)\,dt^2-\frac{1}{r^2\,(1-2\,m/r)^2}\,\Big(q^2+2\,n^2\,\big(\frac{m}{r}-1\big)-r^2\,\big(1-\frac{2\,m}{r}\big)\\
		&\times\big[1+\nu\,\ln\big(1+\frac{m^2}{r^2\,(1-2\,m/r)}\big)\big]\Big)\,dr^2+\Big(n^2+r^2\,\big[1+\nu\,\ln\big(1+\frac{m^2}{r^2\,(1-2\,m/r)}\big)\big]\Big)\,d\theta^2\\
		&+\big(r^2+n^2\big)\,d\phi^2.
	\end{aligned}
\end{equation}
Using the light ring method in the eikonal limit, we obtain the QNMs of the charged Taub-NUT metrics in the presence of a scalar field as follows
\begin{equation}\label{qnmsctns2}
	\begin{aligned}
		Q & = \Big(\Omega + i \, \tilde{\Gamma}\Big)\\
		&=j \, \big[ \frac{1}{3\,\sqrt{3}\,m}\,\big(1\mp\frac{5\,n^2}{18\,m^2}\pm\frac{q^2}{6\,m^2}\big) \big] + i \, \big( n + \frac{1}{2} \big) \, \Big\{ \frac{1}{3 \, \sqrt{3} \, m} \, \big[ 1-\frac{\nu}{2}\,\ln\frac{4}{3}-\frac{4\,n^2}{27\,m^2}+\frac{q^2}{9\,m^2} \big]\Big\}.
	\end{aligned}
\end{equation}
According to Eq. \eqref{qnmsctns2}, the presence of the NUT parameter either increases or decreases the real part of the QNMs, depending on the sign of the relation. Additionally, the NUT parameter reduces the imaginary part of the QNMs.
\section{Conclusion}\label{6}
We derived a class of axially symmetric metrics in the presence of a massless scalar field and incorporated dyonic charge using Harrison transformations. We then obtained the curvature singularities of these metrics. Next, we investigated the astrophysical properties of this class of metrics by deriving the effective potential and studying the motion of neutral particles in the spacetime. Using Ehlers transformations, we added the NUT parameter to the metrics. We also explored gravitational lensing, deriving the deflection angle of light for the metrics with dyonic charge and the NUT parameter in the presence of a massless scalar field. Finally, we calculated the QNMs of these metrics for specific parameter values. There are several interesting avenues for future research, including the investigation of the stability of these metrics and deriving their rotating form. Our exact solutions can also serve as a model for boson stars in the presence of a massless scalar field.
	\section*{Acknowledgements}
	We would like to thank Mahnaz Tavakoli Kachi for her valuable comments. We also acknowledge the financial support provided by Isfahan University of Technology.
	\bibliography{references}{}
	
\end{document}